\documentclass{article}
\usepackage{graphicx} % Required for inserting images
\usepackage{authblk}

\title{Effective potentials for de Sitter and anti de Sitter  quantum fields}
\author{Alfio Bonanno,$^1$ Sergio Luigi Cacciatori$^{2,3}$ and Ugo Moschella$^2$ }

\affil{$^1$ INAF Osservatorio Astrofisico di Catania, Via S.Sofia 78, 95123 Catania Italy, and INFN, Sezione di Catania, Italy}
\affil{$^2$ Department of Science and High Technology, Universit\`a dell'Insubria, Via Valleggio 11, IT-22100 Como, Italy, and INFN sezione di Milano, via Celoria 16, IT-20133 Milano, Italy}
\affil{$^3$ Como Lake centre for AstroPhysics (CLAP), DiSAT, Universit\`a dell’Insubria, via Valleggio 11, 22100 Como, Italy}

\usepackage{color}
\usepackage{amssymb}
\usepackage{amsmath}
\def\phic{\varphi}
\def\phiR{\phi_R}

\def\EE{{\cal E}_d}

\def\E{{\bf R}^N}
\def\bC{{\bf C}}
\def\bR{{\bf R}}

\def\Re{\mathop{\rm Re}\nolimits}

\def\ch{\mathop{\rm ch}\nolimits}
\def\sh{\mathop{\rm sh}\nolimits}

\def\TT{{\cal T}}

\def\interior#1{\setbox1=\hbox{$#1$}\rlap{$#1$}\kern0.4\wd1\raise1.1\ht1%
\hbox{$\scriptstyle \circ$}}

\def\boxit#1#2{\setbox1=\hbox{\kern#1{#2}\kern#1}%
\dimen1=\ht1 \advance \dimen1 by #1 \dimen2=\dp1 \advance \dimen2 by #1
\setbox1=\hbox{\vrule height\dimen1 depth\dimen2\box1\vrule}%
\setbox1=\vbox{\hrule\box1\hrule}%
\advance \dimen1 by .4pt \ht1=\dimen1 \advance \dimen2 by .4pt \dp1=\dimen2
\box1\relax}
\def\endprf{\raise .5ex\hbox{\boxit{2pt}{\ }}}
\def\ifundefined#1{\expandafter\ifx\csname#1\endcsname\relax}

\def\beq{\begin{equation}}
\def\endq{\end{equation}}
\def\beqa{\begin{eqnarray}}
\def\endqa{\end{eqnarray}}

\let\UnmodifSec=\section
\renewcommand{\section}{\setcounter{equation}{0}\UnmodifSec}

\def\bbeta{{\rho}}
\def\aalpha{{\mu}}

\def\bC{{\bf C}}
\def\bR{{\bf R}}

\def\Re{\mathop{\rm Re}\nolimits}

\def\ch{\mathop{\rm ch}\nolimits}
\def\sh{\mathop{\rm sh}\nolimits}

\def\TT{{\cal T}}

\def\interior#1{\setbox1=\hbox{$#1$}\rlap{$#1$}\kern0.4\wd1\raise1.1\ht1%
\hbox{$\scriptstyle \circ$}}

\def\boxit#1#2{\setbox1=\hbox{\kern#1{#2}\kern#1}%
\dimen1=\ht1 \advance \dimen1 by #1 \dimen2=\dp1 \advance \dimen2 by #1
\setbox1=\hbox{\vrule height\dimen1 depth\dimen2\box1\vrule}%
\setbox1=\vbox{\hrule\box1\hrule}%
\advance \dimen1 by .4pt \ht1=\dimen1 \advance \dimen2 by .4pt \dp1=\dimen2
\box1\relax}
\def\endprf{\raise .5ex\hbox{\boxit{2pt}{\ }}}
\def\ifundefined#1{\expandafter\ifx\csname#1\endcsname\relax}

\def\beq{\begin{equation}}
\def\endq{\end{equation}}
\def\beqa{\begin{eqnarray}}
\def\endqa{\end{eqnarray}}

\def\P{{\bf P}}

\renewcommand{\cosh}{\ch}\renewcommand{\sinh}{\sh}

\usepackage{graphicx}  
   
\allowdisplaybreaks[1]

\begin{document}

\maketitle
\abstract{We derive a systematic treatment of one-loop effective potentials for interacting scalar fields in curved spacetimes, providing a general formula valid in arbitrary geometries and explicit results for de Sitter and anti-de Sitter backgrounds. We then compute the effective potential for a scalar $O(N)$ theory on a de Sitter space in any integer dimension. In $d=3$ and dimensional regularization, we extend the calculation up to two loops and compute the $\beta$-function and the anomalous mass dimension. They coincide exactly with flat-space results, despite dramatic curvature modifications to physical masses/couplings. The flat limit $R\to\infty$ recovers Coleman-Weinberg, confirming consistency. Working in $d=3$ dimensions, we repeat the calculation for $AdS_3$ by using point-splitting regularization, obtaining analogous results for the $\beta$-function and anomalous mass dimension. \footnote{Dedicated to Jean Pierre Gazeau on his LXXX Birthday.}}

\section{Introduction}

Effective potentials in quantum field theory (QFT) are a central tool to incorporate quantum corrections into classical potentials, allowing precise analyses of vacuum stability~\cite{Sher,Branchina}, radiative symmetry breaking (as in the Coleman--Weinberg mechanism~\cite{ColemanW}), phase transitions and critical phenomena~\cite{Guida,Kleinert,Zinn-Justin:2002ecy}, as well as applications such as electroweak metastability~\cite{Isidori} or Higgs-driven inflation~\cite{Fairbairn}. In this framework, quantum loop effects are encoded in a field-dependent potential whose minima identify the true quantum vacua, and which provides a compact way of organizing radiative corrections in situations where constant or slowly varying background fields play a distinguished role. 

In flat Minkowski space, the diagrammatic structure and renormalization of the effective potential are by now standard material, and efficient techniques exist to compute it to high loop order and to resum large logarithms in a controlled way.
But realistic phenomena typically unfold in gravitational backgrounds and therefore it is desirable to generalize effective potentials from flat Minkowski space to curved spacetimes, \cite{Anderson, Elizalde, Buchbinder}. Situations of interest include inflationary dynamics in de Sitter (dS) space, quantum fields around black holes, and physics near cosmological horizons, where curvature effects qualitatively modify quantum fluctuations and can render flat-space approximations unreliable. 

De Sitter space in particular plays a distinguished role both as an approximate description of the inflationary epoch \cite{Guth} and as a model of the present accelerated universe, while anti-de Sitter (AdS) space is central to holographic dualities and strongly coupled systems \cite{Wadscft}. In all these contexts, a consistent treatment of the effective potential in the presence of curvature is needed to address questions of vacuum structure, symmetry-breaking patterns, and phase transitions in a genuinely gravitational setting, and to assess, for instance, the stability of metastable vacua in the early universe or in curved backgrounds relevant for high-energy physics.

Extending the effective potential to curved backgrounds entails both technical and conceptual difficulties. In curved spacetime, the notions of vacuum and particle states become considerably more subtle, particularly in the absence of a global timelike Killing vector, as in the de Sitter case, and even more so when the manifold fails to be globally hyperbolic, as occurs for AdS.
Fortunately, in both de Sitter and anti-de Sitter spacetimes, one can adopt a formulation based on maximally analytic two-point functions defined on their complexified manifolds. This framework effectively replaces the usual spectral condition--which cannot be consistently implemented in de Sitter quantum field theory--as well as the standard boundary conditions imposed on modes in the anti-de Sitter setting \cite{bgm,bm,bemgen,umads}. Our analysis will rely heavily on this approach.

In curved spacetime, loop integrals involve Feynman or Schwinger functions defined on non-trivial manifolds and must be regularized and renormalized in a manner consistent with the underlying geometry and its symmetries. The presence of curvature introduces additional couplings, such as the non-minimal interaction term $\xi R \varphi^2$, and leads to a non-trivial interplay between mass parameters, curvature couplings, and the cosmological constant.

In many practical applications, one relies on \emph{ad hoc} extensions of flat-space expressions. However, a systematic derivation valid for general curved backgrounds is less commonly developed. It is therefore desirable to construct a formulation of the effective potential that remains as close as possible to standard flat-space diagrammatic techniques, while being consistently adapted to curved manifolds such as spherical and hyperbolic spaces.

%In recent years, significant progress has been made in understanding loop effects for scalar fields in de Sitter and anti--de Sitter geometries. On the one hand, explicit constructions of two-point functions and propagators on dS and AdS have been developed in the language of ambient space and special functions, with a precise characterization of principal and complementary series representations and their relation to mass parameters \cite{bgm,bm}. On the other hand, detailed studies of loop integrals, such as ``banana'' diagrams in de Sitter and anti-de Sitter spaces, have become available, providing a solid basis for multi-loop calculations on maximally symmetric backgrounds \cite{CEM2024,HSUAdS,bemgen}. Parallel to these developments, there has been sustained interest in computing effective potentials for realistic models (including the Standard Model) in curved spacetimes, with applications to vacuum stability in an inflationary background and to curvature-induced corrections to Higgs physics \cite{Markkanen:2018bfx,Serreau:2011fu}. Nevertheless, most of the existing literature focuses either on one-loop results or on special limits, and less attention has been devoted to a general diagrammatic formulation of the effective potential that is tailored to curved manifolds and can be pushed systematically beyond one loop.

The purpose of this work is twofold. First, we derive a general expression for the one-loop effective potential valid in arbitrary Euclidean curved spacetimes, under a mild condition on the Laplacian acting on two-edge-connected diagrams. Starting from the Coleman--Weinberg observation that the effective potential admits a diagrammatic expansion as the sum of all one-particle-irreducible (1PI) graphs with vanishing external momenta \cite{ColemanW}, we show that polygon vacuum diagrams constructed from Schwinger propagators can be systematically reduced to derivatives of the tadpole diagram with respect to the bare mass squared. This procedure leads to curved-space analogues of relations previously identified in flat space by Lee and Sciaccaluga \cite{Sciaccaluga}, and culminates in a compact equation expressing the derivative of the effective potential with respect to the field-dependent mass in terms of the coincident propagator. We prove that this relation holds on any Euclidean background for which a simple Laplacian identity for two-edge-connected diagrams is satisfied. Our primary examples include the Euclidean de Sitter sphere $S_d$, the Lobachevsky (Euclidean AdS) manifold $H_d$, and their common flat limit, Euclidean space $E_d$.

Second, we apply this general framework to a concrete and physically rich class of models: real scalar fields with quartic self-interaction on de Sitter and anti-de Sitter backgrounds, with particular emphasis on three dimensions. For the $O(N)$ model on the Euclidean de Sitter sphere $S_d$, we employ the maximally analytic two-point function \cite{bgm,bm} and its restriction to the sphere to compute the dimensionally regularized tadpole and, via our general formula, the one-loop effective potential in arbitrary spacetime dimension.

We then focus on $d=3$, where the model is super-renormalizable and directly relevant to statistical and condensed-matter physics, since lower-dimensional quantum field theories often provide effective descriptions of critical phenomena and universality classes \cite{Guida,Kleinert,Zinn-Justin:2002ecy}. In this three-dimensional de Sitter background, we go beyond one loop and compute the effective potential at two loops, combining our tadpole-based method with the explicit expressions for banana integrals recently obtained in de Sitter space \cite{CEM2024}.

This procedure yields a fully renormalized two-loop effective potential in ${ dS}_3$ and allows us to clarify in detail how mass and cosmological-constant renormalization operate in this geometry. In particular, we show that the soft divergences characteristic of super-renormalizable theories are absorbed entirely by mass renormalization, while finite curvature-dependent contributions modify the effective cosmological constant and may be interpreted as radiative shifts in the gravitational sector.

In this case, we worked with the dimensional regularization scheme, where, in $d=3$, only logarithmic divergences appear, from which the renormalization flux and the anomalous mass dimension can be directly read. However, it is remarkable to notice that if one works at $d=3$ with a cut-off regularization, then, the direct calculation of the effective potential up to two loops becomes almost elementary, with the small price of introducing non-logarithmic divergences already at one loop level. 
In order to illustrate this fact, we turn to the analogous $O(N)$ model on ${AdS}_3$, where we compute the effective potential up to two loops using point-splitting regularization rather than dimensional regularization. In this setting the Euclidean propagator admits a particularly simple representation, which enables an explicit evaluation of the two-loop ``watermelon'' diagrams in terms of elementary functions and hypergeometric integrals. The computation exploits a K\"all\'en-Lehmann-type spectral decomposition on the Lobachevsky manifold \cite{bemgen,bemadskl,HSUAdS}, rendering the relevant convolution integrals tractable.

While the renormalization scheme and intermediate expressions differ substantially from the de Sitter case, the final renormalization-group data are remarkably robust. In particular, the anomalous mass dimension and the beta function for a suitably defined dimensionless coupling coincide with their flat-space counterparts, in agreement with general expectations for super-renormalizable theories in three dimensions \cite{Rajantie:1996}.

By contrast, the relation between renormalized parameters $(m,c)$ and physical quantities, such as the pole mass and physical coupling $(m_{\text{phys}},c_{\text{phys}})$, as well as the cosmological constant, is nontrivially affected by curvature in both dS and AdS. In the weak-curvature limit we explicitly recover the known flat-space two-loop effective potential in $d=3$ \cite{Rajantie:1996}. Furthermore, our analysis cleanly disentangles the purely kinematical mass renormalization from genuine curvature-induced contributions, which effectively generate a non-minimal coupling to the background geometry.

From a conceptual standpoint, our analysis highlights how curvature enters the effective potential in multiple ways. In de Sitter space, the same quantum fluctuations that renormalize the mass and quartic coupling also induce finite corrections to the cosmological constant, which can be interpreted as radiative shifts to Newton's constant when the de Sitter radius is treated as a renormalized parameter. In three dimensions, where logarithmic divergences are absent at one loop but appear at two loops, these corrections translate into a curvature-dependent separation between the ``kinematical'' mass and the effective nonminimal $\xi R \phi^2$ coupling; in the weak-gravity regime this separation can be made explicit, and one can track how the physical mass receives both flat-space and curvature-induced contributions. In ${ AdS}_3$, similar considerations apply, although the ultraviolet structure is organized in terms of the point-splitting scale rather than the dimensional regulator, and the spectrum of fluctuations reflects the different global geometry and boundary conditions.

Finally, our results fit naturally into the broader program of using lower-dimensional QFTs as laboratories for critical phenomena and for quantum fields in curved space. Three-dimensional scalar theories with $O(N)$ symmetry provide prototypical examples for universality classes relevant to statistical systems near criticality \cite{Guida,Kleinert,Zinn-Justin:2002ecy}, and the inclusion of de Sitter or anti-de Sitter curvature opens the way to controlled studies of how gravitational backgrounds distort or shift phase structure and critical behaviour. The explicit two-loop effective potentials we obtain in ${\rm dS}_3$ and ${\rm AdS}_3$ offer a concrete starting point for such investigations, including radiative symmetry breaking, metastability, and curvature-induced transitions. They may also be useful in future work on holographic interpretations of scalar effective actions in AdS and on stochastic or infrared approaches to scalar fields in de Sitter space.

The paper is organized as follows. In Section \ref{2}, we revisit the diagrammatic expansion of the effective potential and derive a general formula relating polygon vacuum diagrams to derivatives of the tadpole on curved Euclidean backgrounds satisfying a simple Laplacian identity. This leads to a compact expression for the one-loop effective potential in terms of the coincident Schwinger function and generalizes the Lee-Sciaccaluga equation to curved space. In Section \ref{3}, we apply this framework to the $O(N)$ model on the Euclidean de Sitter sphere, compute the one-loop effective potential in arbitrary dimension, and carry out a full two-loop calculation in three dimensions, including a detailed discussion of renormalization, physical parameters, and the flat limit. In section \ref{4}, we perform the analogous analysis for $AdS_3$, using point-splitting regularization to evaluate the relevant loop integrals and to obtain the two-loop effective potential and renormalization conditions.

\section{The 1-loop effective potential in curved spacetime: a  general formula} \label{2}

In 1973 Coleman and Weinberg \cite{ColemanW} pointed out  that there exists a diagrammatic expansion for
the effective potential: it is the sum of all 1PI graphs with vanishing external momenta. 
At one loop, for a scalar field with quartic self-interaction,  this amounts to the sum of all polygonal diagrams built with the free Feynman propagator:
\begin{eqnarray}
V(\phic)=  \frac 12 m^2 \phic^2+ \frac{c }{4} \phic^4 +V_0(\phic) =  \frac 12 m^2 \phic^2+\frac{c }{4} \phic^4 + i \sum_{n=1}^{\infty}\frac{{\left(  3c  \phic^2\right)^n}}{2n}  \int  \frac{d^4k}{(2\pi)^4} \frac{1}{(k^2-m^2+i\epsilon)^{n}}\, ;
\label{cole}
\end{eqnarray}
here we let all counterterms aside. 

In their seminal paper Coleman and Weinberg discussed the case  where the bare mass $m$ is zero, but the presence of a non-zero mass $m$, as in the above formula, is helpful. Let us consider indeed   the tadpole diagram, which formally amounts to the value at coinciding points of the propagator: 
\begin{eqnarray}
G(0)=\int  \frac{d^4k}{(2\pi)^4} \frac{1}{(k^2-m^2+i\epsilon)};
\end{eqnarray}
by using the elementary formal identity
\begin{eqnarray}
\int  \frac{d^4k}{(2\pi)^4} \frac{1}{(k^2-m^2+i\epsilon)^{n+1}} =  \frac{1}{n!}\  \left(\frac\partial{\partial m^2}\right)^n G(0)
\label{cole2}
\end{eqnarray}
it is possible to reshape Eq. (\ref{cole})  as follows:
\begin{eqnarray}
V(\phic) = \frac 12 m^2 \phic^2+ \frac{c }{4} \phic^4 + i \sum_{n=1}^{\infty}\frac{{\left(  3c  \phic^2\right)^n}}{2n}  \frac{1}{(n-1)!}\  \frac{\partial^{n-1}}{(\partial m^2)^{n-1}} G(0).
\label{cole3}
\end{eqnarray}
We aim to demonstrate that Eq.~(\ref{cole3}) enjoys significantly broader applicability beyond flat spacetime. Our primary examples comprise the de~Sitter sphere $S_d$, the Lobachevsky (Euclidean AdS) manifold $H_d$, and their shared flat limit, the Euclidean space $E_d$; the framework remains however more general.

Consider a two-edge connected diagram featuring two distinct external vertices in a $d$-dimensional Euclidean curved manifold $\EE$:
\begin{eqnarray}
F_{m_1 m_2}(x,y)=
 \int_{\EE} S^{}_{m_1}(x, z)  S^{}_{m_2}(y, z) \sqrt {g(z)}\, dz,  \label{diagram}
\end{eqnarray}
where $S_{m_i}(x,y)$ denote the corresponding Schwinger propagators, with distinct masses $m_1 \neq m_2$.
The distribution $F_{m_1 m_2}(x,y)$ is a joint solution of the following equations: 
\begin{eqnarray}
&& (-\nabla^2_x+m^2_1)F_{m_1 m_2}(x,y)=
 \int \delta (z, x)  S^{}_{m_2}(z, y) \sqrt {g(z)} \,dz = S^{}_{m_2}(x, y), \label{011}
\\
&& (-\nabla^2_y+m^2_{2})F_{m_1 m_2}(x,y)=
 \int S^{}_{m_1}(x\cdot z)  \delta(y, z)  \sqrt {g(z)}\,dz = S^{}_{m_1}(x\cdot y). \label{012}
\end{eqnarray}
Suppose that
\begin{eqnarray}
\nabla^2_x F_{m_1 m_2}(x,y)=\nabla^2_y F_{m_1 m_2}(x,y). \label{conditio}
\end{eqnarray}
This condition holds in particular in de Sitter, anti-de Sitter, and flat spacetimes.

Subtracting Eq.~(\ref{012}) from Eq.~(\ref{011}) shows that the above two-edge diagram reduces to a linear combination of free propagators:
\begin{eqnarray}
F_{m_1 m_2}(x,y) % \int_{{\cal E}d} S{m_1}(x, z) S_{m_2}(z, y) \sqrt{g(z)}, dz 
= -\frac{S_{m_1}(x, y)}{m_1^2 - m_2^2} - \frac{S_{m_2}(x, y)}{m_2^2 - m_1^2}. \label{formu2}
\end{eqnarray}
In the limit $m_1 \to m_2$, this expression becomes the derivative of the propagator with respect to $m^2$:
\begin{eqnarray}
&& F^{}_{m(2)}(x,y)= -\frac {\partial} {\partial m^2} {S^{}_{m}(x, y).}
\end{eqnarray}
 By taking in Eq. (\ref{formu2}) the limit $y\to x$ we deduce a general formula for the bubble vacuum diagram:  
 \begin{eqnarray}
bubble(m_1,m_2) = \int_{{\cal E}_d} S^{}_{m_1}(x, z)  S^{}_{m_2}(z,x) \sqrt {g(z)}\, dz= - \frac{S^{}_{m_1}(x, x) - S^{}_{m_2}(x, x)}{m_1^2-m_2^2}. \label{formu3}
\end{eqnarray}
In particular, for equal masses \
 \begin{eqnarray}
bubble(m) = -\frac {\partial} {\partial m^2} S_{m}(x, x)=  -\frac {\partial} {\partial m^2} \,tadpole(m). \label{formu4b}
\end{eqnarray} 
Iterating the above construction, always supposing the validity  of Eq. (\ref{conditio}),  we compute the  3-edge diagram with two internal vertices as follows:
 \begin{eqnarray}  &&F_{m_1 m_2 m_3}(x,y)= \int_{\EE\times \EE} S^{}_{m_1}(x, x_1)  S^{}_{m_2}(x_1, x_2) S^{}_{m_3}(x_2, y) \sqrt {g(x_1)}\, dx_1 \sqrt {g(x_2)}\, dx_2  
 %\cr &&  = -\int_{{\cal E}_d} \frac{S^{}_{m_1}(x, x_1) - S^{}_{m_2}(x, x_1)}{m_1^2-m_2^2} S^{}_{m_3}(x_1, y) \sqrt {g(x_1)}\, dx_1 =% \cr && =\frac{G^{({\cal E})}_{m_1}(x, y) - G^{({\cal E})}_{m_3}(x, y)}{(m_1^2-m_2^2)(m_1^2-m_3^2)}- \frac{G^{({\cal E})}_{m_2}(x, y) - G^{({\cal E})}_{m_3}(x, y)}{(m_1^2-m_2^2)(m_2^2-m_3^2)}= 
 \cr &&\qquad= \frac{S^{}_{m_1}(x, y)}{(m_1^2-m_2^2)(m_1^2-m_3^2)}+\frac{S^{}_{m_2}(x, y)}{(m_2^2-m_3^2)(m_2^2-m_1^2)}+\frac{S^{}_{m_3}(x, y)}{(m_3^2-m_1^2)(m_3^2-m_2^2)}\label{formu4}\, ;
  \end{eqnarray} 
for equal masses this reduces to 
  \begin{eqnarray}
 F_{m (3)} (x,y)=\frac 1 2 \left( \frac{\partial }
{ {\partial {m^2}} } \right)^2 S^{}_{m}(x, y).
\end{eqnarray}
The limit $y\to x$ gives the triangle  diagrams with vanishing external momenta. In particular, when the three masses are equal 
 \begin{eqnarray}
triangle(m) = \frac 12  \left( \frac{\partial }
{ {\partial {m^2}} } \right)^2\, tadpole(m).
\end{eqnarray}
In general, we may compute the  $(n+1)$-edge diagram with $n$-internal vertices and get the following linear combination of propagators:
\begin{align}  
F_{m_1 m_2,\ldots, m_n}(x,y)&= \int_{{\cal E}_d} S^{}_{m_1}(x, x_1)  \ldots S^{}_{m_3}(x_n, y) \sqrt {g(x_1)}\, dx_1 \ldots \sqrt {g(x_n)}\, dx_n  
 \cr &=( -1)^{n} \sum_{j=1}^n
 \frac{S^{}_{m_j}(x, y)}{\prod_{i\not = j}(m_j^2-m_i^2)}. \label{uneq}
\end{align}
If the masses circulating in the diagram are all equal this formula  reduces to 
\begin{eqnarray}
 F_{m (n)} (x,y)=  \frac {(-1)^{n}} {n!}  \left( \frac{\partial }
{ {\partial {m^2}} } \right)^n  S^{}_{m}(x, y).
\end{eqnarray}
The polygon vacuum diagram with $(n+1)$ propagators is obtained by taking the limit $y\to x$ and is  proportional to the  $n$-th derivative of the tadpole w.r.t. the bare mass squared:
 \begin{eqnarray}
polygon_{n+1}(m) =  \frac {(-1)^{n}}  {n!}  \left( \frac{\partial }
{ {\partial {m^2}} } \right)^n\, tadpole(m). \label{tadn}
\end{eqnarray}
This formula provides a generalization of Eq.  (\ref{cole2}) and  is a consequence of the more general Eq. (\ref{uneq}).   Eqs.  (\ref{uneq})  and (\ref{tadn})   are true on any curved background with Euclidean signature\footnote{Similar formulae exists also  for the chronological propagator in Lorentzian signature.}  provided the identity (\ref{conditio}) holds; in particular they hold for the  flat, spherical, and hyperbolic Euclidean geometries. 
The coincident-vertex limit is typically divergent, requiring regularization and renormalization; in the following we will explore both the dimensional and the UV cutoff regularizations.

Let us now consider  the 1-loop correction in Eq. (\ref{cole3})
\begin{eqnarray}
V_{0}(\phic) =   \sum_{n=1}^{\infty} \frac{1}{2n (n-1)!}\left({3c \phic^{2} } \right)^n \left( \frac{\partial }
{ {\partial {m^2}} } \right)^{n-1} S^{}_{m}(x, x) \label{V1} 
\end{eqnarray}
and take the derivative of both sides with respect to $\phic$:
\begin{eqnarray}
\partial_{\phic} V_{0}(\phic) =   {3 c \phic}  \sum_{n=1}^{\infty} \frac{1}{(n-1)!}\left(3 c  \phic^{2} \right)^{n-1} \left( \frac{\partial }
{ {\partial {m^2}} } \right)^{n-1}  S^{}_{m}(x, x)= {3 c \phic}  S^{}_{M(\phic)}(x, x), \label{LS}
\end{eqnarray}
where we have defined
\begin{equation}
    M^2(\phic)=m^2+{3 c  \phic^2 }.
\end{equation} 
It is useful to rewrite Eq. (\ref{LS}) as follows:
\begin{eqnarray}\frac{\partial V_{0}(\phic)}{\partial{M^2(\phic)}} = %\frac {1}{6c\phi} \partial_{\phic} V_{0}(\phic)  = 
\frac 12  S^{}_{M(\phic)}(x, x). \label{LS2}\end{eqnarray}
In flat space, this is nothing but the Lee–Sciaccaluga equation for the effective potential \cite{Sciaccaluga}. Here, its validity has been established on an arbitrary curved background, provided that condition (\ref{conditio}) is satisfied, and widely generalizes the results in \cite{Inami}, valid in homogeneous and globally static space-times. Moreover, the equation continues to hold to all orders in the loop expansion.
%\footnote{$\lambda \phic$ in our notations is $f$ in their ones - cfr. Eq (3.1) of  \cite{Sciaccaluga}.}.

%Eq. (\ref{V1}) has also a practical value in computing the effective potential directly in position space. 
\vskip 5 pt
Let us  show for instance how to recover the standard result  of Coleman and Weinberg at 1-loop with the above formulae. The Schwinger propagator in flat space behaves at short distances  as follows:
\begin{align}
S_m^{}(x,y) =   
 \frac 1 {(2\pi)^{\frac d 2 }}   \left(\frac{r }{ m}\right)^{1-\frac{d}{2}}  K_{\frac{d}{2}-1}\left( m r \right)
\simeq \frac{r^{2-d}}{4 \pi ^{\frac d 2}   } 
  \Gamma \left(\frac{d}{2}-1\right)+ \frac{m^{d-2}}{(4 \pi) ^{\frac d 2}}  \Gamma \left(1-\frac{d}{2}\right) \label{sdb}
\end{align} 
where $r = |x-y|$.  For $d < 2$, only the second term survives at $r = 0$, yielding
\begin{eqnarray}
\left(\frac{\partial}{\partial m^2}\right)^{n-1} S_m(0) = (-1)^{n-1} \frac{(m^2)^{\frac d 2 - n}}{(4\pi)^{\frac d 2}} \Gamma\left( n - \frac{d}{2} \right).
\end{eqnarray}
Inserting this expression into Eq.~(\ref{V1}) and carrying out the summation, we obtain—after analytic continuation—the one-loop contribution to the effective potential in closed form. The resulting expression is a simple meromorphic function of the spacetime dimension~$d$:
\begin{eqnarray}V_0(\phic)= \frac{1} {2^{1+d} \pi^{\frac d2}}\Gamma \left(-\frac{d}{2}\right)
   \left(m^d-\left(m^2+{3 c  \phic^2 }\right)^{\frac d 2 }\right). \label{225}
   \end{eqnarray}
   Integrating  Eq. (\ref{LS2}) gives obviously the same result provided the arbitrary constant term is fixed by requiring $V_0(0)=0$.
   %\begin{eqnarray}V_0(\phic)= \frac{1} {2^{1+d} \pi^{\frac d2}}\Gamma \left(-\frac{d}{2}\right)\left(C-\left(m^2+{3 c  \phic^2 }\right)^{\frac d 2 }\right). \label{225}\end{eqnarray}
   
   In odd dimensions and Eq. (\ref{225}) is already regular; in particular in $d=1$ and $d=3$ we get 
   \begin{eqnarray}
   &d=1:& \ \ \  V^{(1)}_0(\phic)=  \frac{1}{2} \left(\sqrt{m^2+3 c \phic^2}-m\right), \\
    &d=3:&\ \ \  V^{(3)}_0(\phic) = \frac{m^3-\left(m^2+3 c \phic ^2\right)^{3/2}}{12 \pi }.
   \end{eqnarray}

In even  dimensions and Eq. (\ref{225}) is divergent and needs to be regularized and normalized to get the finite result. Let us  extract formulae in $d=2-2\epsilon$ and $d=4-2\epsilon$. We may drop the constant term and get 
\begin{eqnarray}
 &d=2-2\epsilon:&    V_0^{(2)}(\phic)=\left(\frac{3 (1-\gamma )+\log \left(64 \pi ^3\right)}{8 \pi }+\frac{3}{2 \pi  \epsilon}\right)\phic^2  -\frac{\left(m^2+3c \phic ^2\right) \log \left(m^2+3 c\phic ^2\right)}{8 \pi } 
\cr&& +\, O(\epsilon)
  \\  &d=4-2\epsilon:& V_0^{(4)}(\phic)=%&-\frac{2^{-5+2\epsilon}}{\pi^{2-\epsilon}} \Gamma \left(-2+\epsilon\right) \left(m^2+{3 c  \phic^2 }\right)^{2-\epsilon}\cr
   -\frac 1{64\pi^2} \left(m^2+{3 c \phic^2 }\right)^{2}\left(\frac 1\epsilon +\log (4\pi)-\gamma\right) +\cr
   &&+\frac 1{64\pi^2} \left(m^2+{3 c \phic^2 }\right)^{2}\left( \log \left(m^2+{3 c \phic^2 }\right)-\frac 32 \right)+O(\epsilon).
\end{eqnarray}
In both cases the first term %proportional to $\frac 1\epsilon +\log (4\pi)-\gamma$, where $\gamma$ is the Euler-Mascheroni constant,
contribute to the renormalization of the bare tree-level constants. The remaining term reproduces  the one-loop effective potential in $d=2,4$ for a real  scalar field  with quartic self-interaction \cite{ColemanW, Sciaccaluga}.

\section{The $O(N)$ model on the Euclidean de Sitter  sphere}\label{sec2}
\label{3}

Let us start by briefly recalling some facts about de Sitter (scalar) quantum fields. Let  $M_{d+1}$
be the real
$(d+1)$-dimensional Minkowski space-time and $M_{d+1}^{(c)}$
be its complexification. 
In a chosen Lorentz frame %$\{e_\mu, \ \mu=0,\ldots,d\}$
the scalar product of two (complex) events is  
\begin{equation} z_1\cdot z_2  = z^0_1 z^0_2-z^1_1 z^1_2 - \ldots -z^d_1 z^d_2.
\end{equation}
The (complex) de Sitter  universe  may be represented as  the  one-sheeted hyperboloid immersed in the (complex) Minkowski space $M_{d+1}^{(c)}$:
\beq
dS_d^{(c)} = \{z \in M_{d+1}^{(c)}\ :\ z\cdot z = -R^2\}\ .
\label{s.2}\endq
The de Sitter invariant  complex variable $\zeta$ is the scalar product in the ambient spacetime of  two complex events $z_1, z_2 \in dS_d^{(c)}$:  
\begin{equation}
    \zeta= \frac{z_1\cdot z_2}{R^2}.
\end{equation} 
The future and past tuboids $\TT_\pm$, which encode the spectral condition for quantum field theory on de Sitter space, are defined as the intersections of the ambient forward and backward tubes $T_\pm$ with the complexified de Sitter manifold:
\beq
\TT_\pm = \{ x + i y \in dS_d^{(c)} \; : \; y \in \pm V_+ \} \, .
\label{s.2.1}
\endq
Because de Sitter spacetime does not admit a global timelike Killing vector, a standard spectral condition cannot be formulated. Instead, one imposes the requirement of \emph{normal analyticity}~\cite{bgm,bm}: the two-point distributions must arise as boundary values of functions holomorphic in the domain $\TT_- \times \TT_+$. Combined with de Sitter invariance and the canonical commutation relations, this condition uniquely fixes the two-point function of any de Sitter massive Klein--Gordon field  \cite{bgm,bm,CEM2024}:
\begin{align}
W^d_\nu (z_1,z_2)    &= \frac {\Gamma\left(\frac{d-1}2 + i \nu\right)\Gamma\left(\frac{d-1}2 - i \nu\right)}{2 (2\pi)^{d/2}}
(\zeta^2-1)^{-\frac {d-2}4}\,P_{ -{\frac 12}+ i \nu }^{-\frac {d-2}2}(\zeta) \label{leg2p}
 \\  &= 
\frac {\Gamma\left(\frac{d-1}2 + i \nu\right)\Gamma\left(\frac{d-1}2 - i \nu\right)}{(4\pi)^{d/2}\Gamma\left(\frac d2\right)}
{}_2F_1\left(\frac{d-1}2 + i \nu,\frac{d-1}2 - i \nu;\ {\frac d2};\ \frac {1-\zeta}2\right ).
\label{wig}
\end{align}
The complex parameter $\nu$ is related to the complex mass squared as follows:
\begin{equation}
m^2 =  \frac {(d-1)^2}{4} +\nu^2. \label{cmass}
\end{equation}
The squared mass is real and positive only when: 
\begin{itemize}
\item  $\nu$  is real; this corresponds in a group-theoretical language to the principal series of unitary representations of the Lorentz group; 

\item   $\nu$ is purely imaginary such that $|\nu|<\frac{d-1}{2} $; this corresponds to the complementary series of unitary representations of the Lorentz group.
\end{itemize}
There is also a discrete series of acceptable QFT's with negative squared mass \cite{Tachyons1,Tachyons2}.
Finally, the Schwinger function (in short: the propagator) is the restriction of the maximally analytic two-point function to the Euclidean sphere
\beq
S_d^{} = \{z_{\scriptscriptstyle{E}} \in dS_d^{(c)},\   z_{\scriptscriptstyle{E}}^0=i x^0, \  z_{\scriptscriptstyle{E}}^{i} =x^i,\  x^{\mu}\in {\bf R},\  -R\leq x^\mu\leq R \}
\label{s.2};\endq
in this case the scalar product may be parametrized by an angle $s$ so that   ${z_{\scriptscriptstyle{E}}}_1\cdot 
{z_{\scriptscriptstyle{E}}}_2 = -R^2\cos s$  and 
\begin{eqnarray}
 G^d_{\nu}(- \cos s ) = \frac {\Gamma({\frac {d-1}{2} + i\nu} )\Gamma({\frac {d-1}{2} - i\nu} )}{2 (2\pi)^{d/2}} (\sin s)^{-\frac {d-2}{2}} \P_{-\frac 12 +i \nu }^{-\frac {d-2}2}(-\cos s)  \label{leg6i}
\end{eqnarray}
where
$\P_\bbeta^\aalpha(z)$
is the so called "Legendre function on the cut"  \cite{bateman} or Ferrers function of the first kind.\footnote{It is important to  keep in mind that  Ferrers  functions $\P_\beta^\alpha(z)$ and Legendre functions $P_\beta^\alpha(z)$   are holomorphic in
different cut-planes;  as regards Ferrers function this is 
\beq
\Delta_2 = \bC\setminus \{(-\infty,-1]\cup [1,\infty)]\}.
\label{d.20}\endq}

\vskip 10 pt 

Let us  consider now  a scalar multi-component field 
\begin{align}
 \phiR: S_d \longrightarrow \E,
\end{align}
 where  $S_d$ is the $d$-dimensional Euclidean de Sitter sphere of radius $R$. % or the Anti-de Sitter Lobachevsky space $M_d$ of radius $R$. %or simply the flat Euclidean space $E_d$.  
%The radius $R$ and the cosmological constant \begin{equation}  \Lambda=\frac{(d-1)(d-2)}{2R^2}\end{equation} 
% For the moment we take $R=1$. 
The action for $\phiR$ is  the quartic $O(N)$-invariant action
\begin{align}
 S[\phi]=\int_{\EE}  \Big[ \Lambda_0+\frac 12 \langle \partial_\mu \phiR \mid \partial^\mu \phiR \rangle +\frac {m_R^2}2 \langle \phiR \mid \phiR \rangle +\frac {c_R}4 \langle \phiR \mid \phiR \rangle^2 \Big] \sqrt {g}\, d^d x,
\end{align}
where the standard  scalar product in $\E $ is denoted by $\langle \cdot\mid \cdot \rangle$; the suffix $R$ means that the quantities are made adimensional by rescalings with the de Sitter radius $R$. So, $m_R=m_0 R,\ \phiR=\phi_0 R^{\frac {d-2}2},\ c_R=c_0 R^{4-d} $; $m_0$ and $c_0$ are the bare mass and bare self-coupling constant, respectively. %$x^\mu$ are local coordinates on $\EE$.
We will compute the effective potential for the constant configuration 
\begin{align}
 \overline {\phiR}\equiv \varphi_R\, e_0,
\end{align}
where $\varphi_R$ is a real constant and $e_0$ is a given vector of norm $1$ in $\E$. Of course a nonzero expectation value of $ \phiR$  breaks the symmetry down to $O(N-1)$. After choosing  any orthonormal basis $\{e_j\}_{j=0}^{N-1}$ in $\E$ whose first element is $e_0$, we may write 
\begin{align}
 \phiR=(\varphi_R+\psi_0)e_0+\sum_{j=1}^{N-1} \psi_j e_j,
\end{align}
so that \cite{CEM2024}
\begin{align}
 S[\phiR]=& \int_{S_d} \Big[ \Lambda_0+ \frac {m_R^2}2 \varphi_R^2 +\frac {c_R}4 \varphi_R^4 \Big]\sqrt {g}\, d^d x +\sum_{a=1}^4 S_a[\psi;\varphi],
\end{align}
where
\begin{align}
 S_1[\psi;\varphi_R]=& \int_{S_d}  \Big[{m_0^2} \varphi_R \psi_0 + {c_R} \varphi_R^3 \psi_0 \Big]\sqrt {g}\, d^d x,\\
 S_2[\psi;\varphi_R]=& \frac 12 \int_{S_d}  \Big[\sum_{j=0}^{N-1}\partial_\mu \psi_j \partial^\mu \psi_j +  M_0^2(\varphi_R) \psi_0^2 +  M_1^2(\varphi_R) \sum_{j=1}^{N-1}\psi_j^2 \Big] \sqrt {g}\,d^d x, \\
 S_3[\psi;\varphi_R]=& \int_{S_d}  \Big[c_R \varphi \psi_0 \sum_{j=0}^{N-1} \psi_j^2 \Big] \, d^d x,\\
 S_4[\psi;\varphi_R]=& \frac {c_R}4\int_{S_d} \sqrt {g} \Big[\sum_{j=0}^{N-1} \psi_j^2 \Big]^2 \sqrt {g}\, d^d x.
\end{align}
For $a=0,1$ we have set
\begin{align}
%& {V}_a(\varphi_R)= - \frac 1{\Omega_d} \log \int [D\Phi] \exp \left(-\frac 12 \int_{M_d} \sqrt {g} \Big[\partial_\mu \Phi \partial^\mu \Phi + M_a^2(\varphi_R) \Phi^2 \Big] d^d x\right), \\
& M_0^2(\varphi_R)= m_R^2+3c_R \varphi_R^2, \\
&  M_1^2(\varphi_R)= m_R^2+c_R \varphi_R^2.
\end{align}
For a constant field configuration, the effective potential ${V}(\varphi_R)$ can be determined as follows \cite{Weinberg}
\begin{align}
 \exp (-\Omega_d {V}(\varphi_R))=\int [\prod_j D\psi_j] \exp (-S[\phiR]);
\end{align}
here $\Omega_d$ is the volume of  the sphere $S_d$ and $[\prod_j D\psi_j]$ is the formal path integral measure.
By construction, $S_1$ does not contribute to the effective potential. 
Dropping it, we can write the complete effective potential as
\begin{align}
 {V}(\varphi_R)=%& \Lambda_0+ \frac {m_R^2}2 \varphi_R^2 +\frac {c_R}4 \varphi_R^4-\frac 1{\Omega_d}\log \left(\int [\prod_j D\psi_j] \exp (-S_2[\phiR]-S_3[\phiR]-S_4[\phiR])\right)\cr=
 &  \Lambda_0+ \frac {m_R^2}2 \varphi_R^2 +\frac {c_R}4 \varphi_R^4-\frac 1{\Omega_d}\log \left(\int [\prod_j D\psi_j] \exp (-S_2[\phiR])\right) \cr &- \frac 1{\Omega_d}\log \frac {\int [\prod_j D\psi_j] \exp (-S_2[\phiR]-S_3[\phiR]-S_4[\phiR])}{\int [\prod_j D\psi_j] \exp (-S_2[\phiR])}. \label{313}
\end{align}
The first line at the rhs is the one loop contribution while the second line is the sum of all connected vacuum diagrams. 

In the following we study the general case at one-loop and the three-dimensional case at two-loop. Our strategy will be to use the tadpole equation deduced in the previous section to compute the one-loop contribution, and the known results on the banana integrals in de Sitter, recently obtained in \cite{CEM2024}.

%%%%%%%%%%%%%%%%%%%%%%%%%%%%%%%%%%%%%%%%%%
\subsection{The 1-loop correction}
The 1-loop effective potential at the rhs of (\ref{313}) can be rewritten as follows:
\begin{align}
 V_{1{\rm loop}}(\varphi_R)=&  \Lambda_0+ \frac {m_R^2}2 \varphi_R^2 +\frac {c_R}4 \varphi_R^4 +{V}_0(\varphi_R)+(N-1) {V}_1(\varphi_R).
\end{align}
%where, for $a=0,1$,
%\begin{align}
%& {V}_a(\varphi_R)= - \frac 1{\Omega_d} \log \int [D\Phi] \exp \left(-\frac 12 \int_{M_d} \sqrt {g} \Big[\partial_\mu \Phi \partial^\mu \Phi + M_a^2(\varphi_R) \Phi^2 \Big] d^d x\right), \\
%& M_0^2(\varphi_R)= m_R^2+3c_R \varphi_R^2, \\
%&  M_1^2(\varphi_R)= m_R^2+c_R \varphi_R^2.
%\end{align}
 The starting point to  solve the tadpole equation (\ref{LS2}) for a massive quantum scalar field is its maximally analytic two-point function (\ref{wig}). 
 In dimensional regularization 
the tadpole may be computed by taking the limit where the two point in Eq. (\ref{wig}) coincide; this limit %i.e. $z_1\cdot z_2/R^2=-1$; 
 is finite  for $\Re d<2$ and we take its meromorphic continuation to any complex dimension $d$:
\begin{align}
T_d(\nu)=%&  W^d_\nu (z,z)    =\frac {\Gamma\left(\frac{d-1}2 + i \nu\right)\Gamma\left(\frac{d-1}2 - i \nu\right)}{(4\pi)^{d/2}\Gamma\left(\frac d2\right)}{}_2F_1\left(\frac{d-1}2 + i \nu,\frac{d-1}2 - i \nu;\ {\frac d2};\  1\right) \cr  =
&  \frac{2^{-d} \pi ^{-d/2} \Gamma \left(1-\frac{d}{2}\right) \Gamma \left(\frac{d-1}{2}+
   i \nu \right) \Gamma \left(\frac{d-1}{2} - i \nu \right)}{\Gamma
   \left(\frac{1}{2}+i \nu \right) \Gamma \left(\frac{1}{2}-i \nu \right)}.
\label{tadds}
\end{align}
 The  meromorphic  continuation in turn provides a dimension-regularized integral representation of the 1-loop correction to the bare potential: 
\begin{eqnarray}
V_0(\phic_R)=2^{-d} \pi ^{-d/2} \Gamma \left(1-\frac{d}{2}\right)\int^{\nu(\phic_R)}    \frac{ \Gamma \left(\frac{d-1}{2}+
   i x\right) \Gamma \left(\frac{d-1}{2} - i x \right)}{\Gamma
   \left(\frac{1}{2}+i x\right) \Gamma \left(\frac{1}{2}-i x \right)}  x dx  \label{325}
\end{eqnarray}
where 
\begin{equation}
    \nu(\phic_R)=\left|m_R^2+3c_R \phic_R^2-\frac{(d-1)^2}{4}\right|^{\frac 12}.
\end{equation}
\subsubsection{Even dimensions}
Let us briefly discuss first the even-dimensional case which demands regularization. We restrict this discussion to $N=1$ and  we suppose that the bare mass is in the principal series. We limit ourselves to  giving the formulae that are easily deduced from Eq. (\ref{tadds}); in dimension $d=4$ the formula first appeared in \cite{CEM2024} and was then found again in \cite{VicenteGarcia-Consuegra:2025two} (see also  \cite{Fujimoto}, \cite{Shore}, \cite{Esposito} for early calculations in $d=4$).
\vskip 5 pt
 $d=2-2\epsilon$. 
    \begin{eqnarray}
     && V_0^{(2)}(\phic_R)=   \sqrt{4 m_R^2+12 c_R \phic_R ^2-1} \left(\frac{1}{8 \pi  \epsilon }+\frac{\log (4 \pi
   )-\gamma }{8 \pi }\right)+\cr&&  +\frac{i \left(
   {\log \Gamma }\left(\frac{1}{2}+i
   \sqrt{m_R^2+3 c_R \phic_R ^2-\frac{1}{4}}\right)\right)-{\log \Gamma }\left(\frac{1}{2}-i
   \sqrt{m_R^2+3 c_R \phic_R ^2-\frac{1}{4}}\right)}{4 \pi } 
    \end{eqnarray}  

$d=4-2\epsilon$. 
Here the dimension-regularized tadpole is given by 
\begin{equation}
    T_{4-2\epsilon}(\nu)= -\frac{\left(1+4 \nu^2\right) \left(\frac{1}{\epsilon }+1-\gamma +\log (4 \pi )\right)}{64
   \pi ^2}
+\frac{\left( \nu^2+\frac 14 \right) \left(\psi \left(-\frac{1}{2}-i \nu\right)+\psi
   \left(-\frac{1}{2}+i \nu\right)\right)}{16 \pi ^2}.
\end{equation}
Integration gives
    \begin{eqnarray}
     && V_0^{(4)}(\phic_R)=  -\frac{\left(x^4+\frac{x^2}{2}\right) \left(\frac{1}{\epsilon }-\gamma +1+\log (4 \pi
   )\right)-\left( \nu^2+\frac 14\right)^2 }{64  \pi ^2} + \cr && +  \frac { \left( x^2+\frac 14\right)^2 }{32\pi^2}
\left.\left(\psi \left(-\frac{1}{2}+ix\right)+\psi
   \left(-\frac{1}{2}-i x\right)\right)  +\frac{B(x)}{16\pi^2}\  \right|_{ x= \sqrt{m_R^2+3c_R \phic_R^2-\frac{9}{4}}
   } \label{3.30}
    \end{eqnarray}
 where 
  %  \begin{eqnarray}  x= \sqrt{m_R^2+3c_R \phic_R^2-\frac{9}{4}}  \end{eqnarray}
%It is useful to rewrite the integral of the second term \cite{} as follows:
%\begin{eqnarray}&& \int {\left( \nu^2+\frac 14 \right) \left(\psi \left(-\frac{1}{2}-i \nu\right)+\psi   \left(-\frac{1}{2}+i \nu\right)\right)} \nu d\nu =   \cr && =\frac { \left( \nu^2+\frac 14\right)^2 }{2}\left(\psi \left(-\frac{1}{2}+i\nu\right)+\psi \left(-\frac{1}{2}-i \nu\right)\right) \frac { \left( \nu^2+\frac 14\right)^2 }{4} +B(\nu) \cr&& 
\begin{eqnarray}
&&B(x)= {\frac { \left( x^2+\frac 14\right)^2 }{4}+ \int^x \frac i2 \left( y^2+\frac 14\right)^2 \left(\psi' \left(-\frac{1}{2}+i y\right)-\psi'
   \left(-\frac{1}{2}-i y\right)\right)}dy= 
 \cr
&&=-\frac 12 \int dy\ y^2 \left(\sum_{n=0}^\infty \frac {2n-1}{(n(n-1)+y)^2} -\frac 1y\right), \label{Bfunction}
\end{eqnarray}
The function $B$ may also be expressed in terms of antiderivatives of the function $\psi$ as follows \cite{VicenteGarcia-Consuegra:2025two}
\begin{eqnarray}
B(x)&=& -\frac{1}{2} i x \left(4 x^2+1\right) \left(\text{log$\Gamma $}\left(-\frac{1}{2}-i
   x\right)-\text{log$\Gamma $}\left(-\frac{1}{2}+i x\right)\right) \cr  &&+ 12 \psi ^{(-4)}\left(-\frac{1}{2}-i
   x\right)+12 \psi ^{(-4)}\left(-\frac{1}{2}+i x\right)\cr  &&+ 12 i x \psi ^{(-3)}\left(-\frac{1}{2}-i
   x\right)-12 i x \psi ^{(-3)}\left(-\frac{1}{2}+i x\right) \cr &&-\frac{1}{2} \left(12 x^2+1\right) \left(\psi ^{(-2)}\left(-\frac{1}{2}-i
   x\right)+\psi
   ^{(-2)}\left(-\frac{1}{2}+i x\right)\right)+\cr&&+\frac{1}{32} \left(4 x^2+1\right)^2 \left(\psi\left(-\frac{1}{2}-i
   x\right)+\psi
   \left(-\frac{1}{2}+i x\right)\right)
\end{eqnarray}
a formula which however is not particularly illuminating; the above integral representation of the function $B$ is actually more useful. 
 \subsubsection{Odd dimensions} 
No further regularization is needed when using Eq. (\ref{325})  to obtain the 1-loop effective potential in odd dimensions. 
The  simplest case is when $d=1$:
\begin{equation}
T_1(\nu) = \frac{\coth (\pi  \nu)}{2 
\nu}
\end{equation}
so that \begin{equation}
 V_0(\phic_R)  % = \int T_1(\nu) \, \nu\, d\nu
 = \frac 1 {2\pi} \log \left(\frac{\sinh\left(\pi  \sqrt{m_R^2+3 c_R \phic_R^2}\right) }{\sinh\left(\pi  \sqrt{m_R^2}\right)}\right).
\end{equation}

\vskip 5 pt
\noindent The situation becomes more delicate at $d=3$. Here the parameter $\nu$ may be real (principal series) or purely imaginary (complementary series). Let us consider at first  $\nu$  real and positive so that 
\begin{align}
 &T_3(\nu) =   - \frac{\nu  \coth (\pi  \nu )}{4 \pi } \label{T3}
\\
 &V_{0p}(\nu) = \int T_3(\nu) \, \nu\, d\nu
 = -\frac{\nu ^3}{12 \pi }-\frac{\nu ^2 \log \left(1-e^{-2 \pi  \nu }\right)}{4 \pi
   ^2}+\frac{\nu \,  \text{Li}_2\left(e^{-2 \pi  \nu }\right)}{4 \pi^3}+\frac{\text{Li}_3\left(e^{-2 \pi  \nu }\right)}{8 \pi ^4}.\label{pot3ds}
\end{align}
For a single field (i.e. $N=1$) the above integral gives the following 1-loop regularized effective   potential valid when $m_R\geq 1$:  \begin{align}&V_{\rm 1loop}(\phic_R)=\Lambda_0+ \frac {m_R^2}2 \varphi_R^2 +\frac {c_R}4 \varphi_R^4+\cr &-\frac{\left(m_R^2+3 c_R \phic_R^2-1\right)^{3/2}}{12 \pi }-\frac{\left(m_R^2+3 c_R \phic_R^2-1\right) \log \left(1-e^{-2 \pi   \sqrt{m_R^2+3 c_R \phic_R^2-1}}\right)}{4 \pi ^2}+\cr& +  \frac{\sqrt{m_R^2+3 c_R \phic_R^2-1} \ \text{Li}_2\left(e^{-2 \pi  \sqrt{m_R^2+3 c_R \phic_R^2-1}}\right)}{4 \pi   ^3}+\frac{\text{Li}_3\left(e^{-2 \pi  \sqrt{m_R^2+3 c_R \phic_R -1}}\right)}{8 \pi ^4}.  \end{align}
Extrema of the potential are located by solving the equation 
\begin{equation}
    \partial_\phic V= -\frac{3 c_R^2 \phic_R  \sqrt{3 c_R^2 \phic_R ^2+m_R^2-1} \coth \left(\pi  \sqrt{3 c_R^2 \phic_R ^2+_Rm^2-1}\right)}{4 \pi }+c_R \phic_R ^3+m_R^2
   \phic_R =0
\end{equation}
which necessarily  has an odd number of solutions.

A bare mass in the complementary series ($m_R<1$) corresponds instead to a purely imaginary $\nu=i a$, with $|a|<1$. In this case the tadpole is given by  
\begin{eqnarray}
T_3(ia)=  -\frac{ a  \cot (\pi  a )}{4 \pi } \label{T3c}.
\end{eqnarray}
The (indefinite) integral  in Eq. (\ref{325})  should now be considered for values between $0<a<1$. Not surprisingly the result is the analytic continuation of Eq. (\ref{pot3ds}):
\begin{align}
 &V_{0c}(a)= \frac 12 \int^a T_3(i x) \, 2ix\, dix=   \int^a \frac{x^2  \cot (\pi  x )}{4 \pi }  dx =\cr &=  
 \frac{i a^3}{12 \pi }+\frac{a ^2 \log \left(1-e^{-2i \pi  a }\right)}{4 \pi
   ^2}+\frac{i a \,  \text{Li}_2\left(e^{-2 i \pi  a }\right)}{4 \pi^3}+\frac{\text{Li}_3\left(e^{-2 i\pi  a }\right)}{8 \pi ^4}.\label{pot3dscomp}
\end{align}
Notwithstanding the appearance the result is a real function of $a$. 
An alternative manifestly real expression for the 1-loop correction may be obtained by taking the  series expansion of the integrand:
\begin{equation}
 \frac{x^2  \cot (\pi  x )}{4 \pi }  = \frac{x}{4 \pi ^2}-\frac 1 {2 \pi ^2}{\sum _{n=1}^{\infty } x^{2 n+1} \zeta (2 n)}.
\end{equation}
Integrating term by term and resumming we get
\begin{eqnarray}
  V_{0c}(a)&=&  -\frac{\zeta
   ^{(1,0)}(-2,1-a)+\zeta ^{(1,0)}(-2,1+a)+2 a \zeta ^{(1,0)}(-1,1-a)-2 a \zeta
   ^{(1,0)}(-1,1+a)}{4 \pi ^2}\cr &&-\frac{a^2 \log \left(\frac{1}{2} a \csc (\pi  a)\right)}{4 \pi ^2},
   \end{eqnarray}
where $\zeta(s,q)$ is the Hurwitz zeta function.

If the starting bare mass  $m_R<1$ belongs to the complementary series  the regularized effective potential is constructed as follows:  
\begin{align}
&V(\phic_R)=\Lambda_0+ \frac {m_R^2}2 \varphi_R^2 +\frac {c_R}4 \varphi_R^4+V_0(\phic_R)
   \end{align}
   where 
   \begin{equation}
   V_0(\phic_R)=\left\{ \begin{array}{cc}
       V_{0c} (\sqrt{1-m_R^2-3 c_R \phic_R^2}),& \phic_R^2 \leq \frac{1-m_R^2}{3 c_R} \\ &\\
       V_{0p} (\sqrt{m_R^2+3 c_R \phic_R^2-1}),  & \phic_R^2 \geq \frac{1-m_R^2}{3 c_R}. 
   \end{array}   \right. 
   \end{equation}
\vskip 10 pt
   
In the general $O(N)$-model there are one field of mass $M_0(\varphi_R)$ and $(N-1)$ fields of mass $M_1(\varphi_R)$. If all the bare parameters are in the principal series we have 
\begin{align}
    &{V}_{\rm 1loop}(\varphi_R)= \Lambda_0+ \frac {m_R^2}2 \varphi_R^2 +\frac {c_R}4 \varphi_R^4\cr
 &-\frac{\nu_{0}^3}{12 \pi }-\frac{\nu_{0}^2 \log \left(1-e^{-2 \pi  \nu_{0} }\right)}{4 \pi^2}+\frac{\nu_{0} \text{Li}_2\left(e^{-2 \pi  \nu_{0} }\right)}{4 \pi^3}+\frac{\text{Li}_3\left(e^{-2 \pi  \nu_{0} }\right)}{8 \pi ^4}\cr
 &+(N-1)\left(-\frac{\nu_{1}^3}{12 \pi }-\frac{\nu_{1}^2 \log \left(1-e^{-2 \pi  \nu_{1} }\right)}{4 \pi^2}+\frac{\nu_{1} \text{Li}_2\left(e^{-2 \pi  \nu_{1} }\right)}{4 \pi^3}+\frac{\text{Li}_3\left(e^{-2 \pi  \nu_{1} }\right)}{8 \pi ^4}\right), \label{pot3}
\end{align}
where 
\begin{equation}
\nu_{0}=\sqrt{M_0^2(\varphi_R)-1} = \sqrt{ m^2_R+3c^2_R \varphi_R^2-1 }, \ \ \ \nu_{1}=\sqrt {M_1^2(\varphi_R)-1}=\sqrt{ m^2_R+c^2_R \varphi_R^2-1 }.
\end{equation}
Eq. (\ref{pot3}) is valid as such if $m_R>1$. Otherwise it has to be understood in the analytic continuation sense as in the $N=1$ case.
%\begin{figure}    \centering\includegraphics[width=0.5\linewidth]{principal.pdf}    \caption{Principal series}    \label{fig:placeholder}\end{figure}
%\begin{figure}   \centering\includegraphics[width=0.5\linewidth]{princi.pdf}   \caption{Principal series: total 0-loop + 1-loop }   \label{fig:placeholder6}  \end{figure}
\begin{figure}
    \centering
\includegraphics[width=0.4\linewidth]{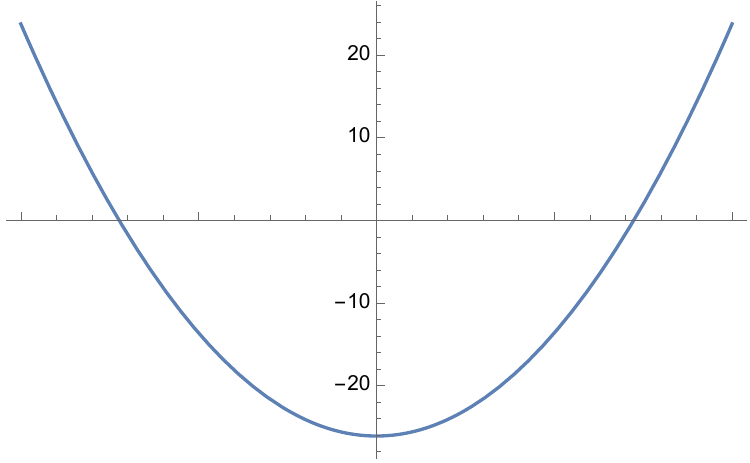} 
\includegraphics[width=0.4\linewidth]{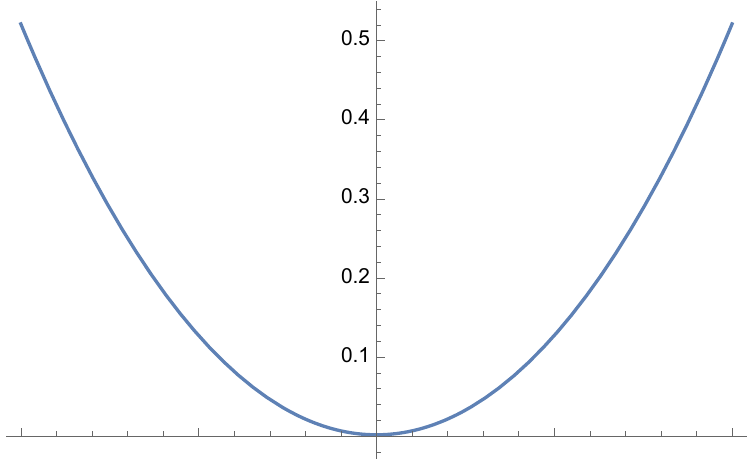} 
\includegraphics[width=0.4\linewidth]{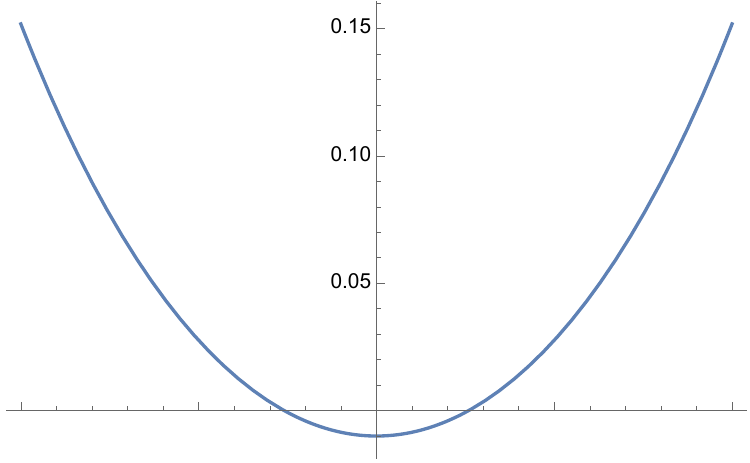} 
\includegraphics[width=0.4\linewidth]{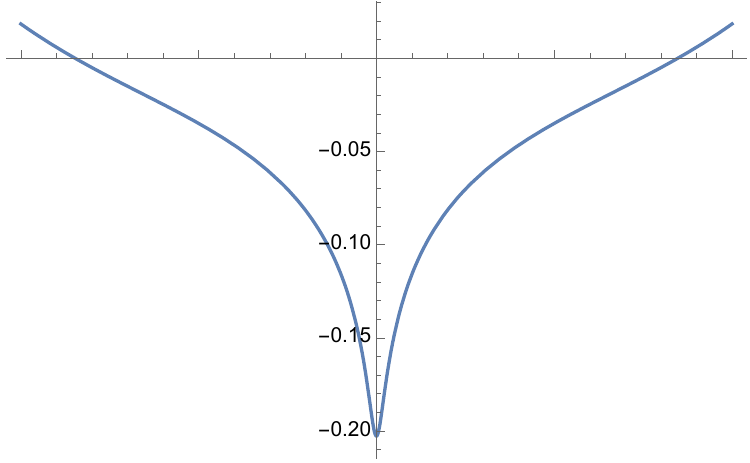} 
    \caption{1-loop effective potential in the de Sitter $d=3$ case. The bare coupling constant is $c_R=0.1$. The masses from left to right: principal series $m_R=10$, limit case $m_R=1$, complementary series $m_R=0.5$ and $m_R=0.001$.}
    \label{fig:1}
\end{figure}
\begin{figure}
    \centering
\includegraphics[width=0.4\linewidth]{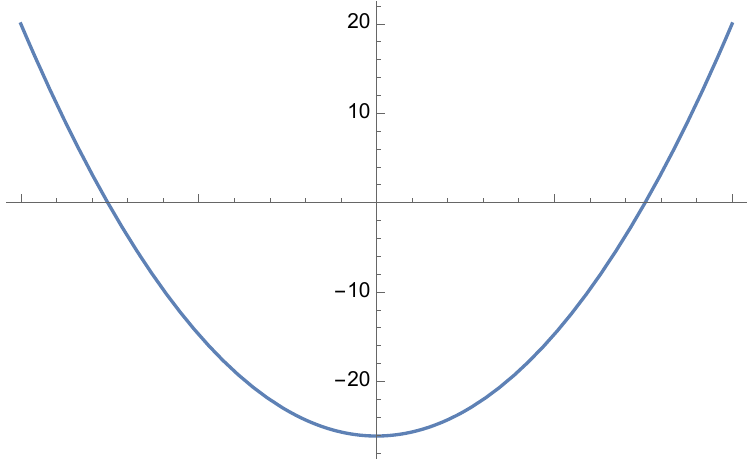} 
\includegraphics[width=0.4\linewidth]{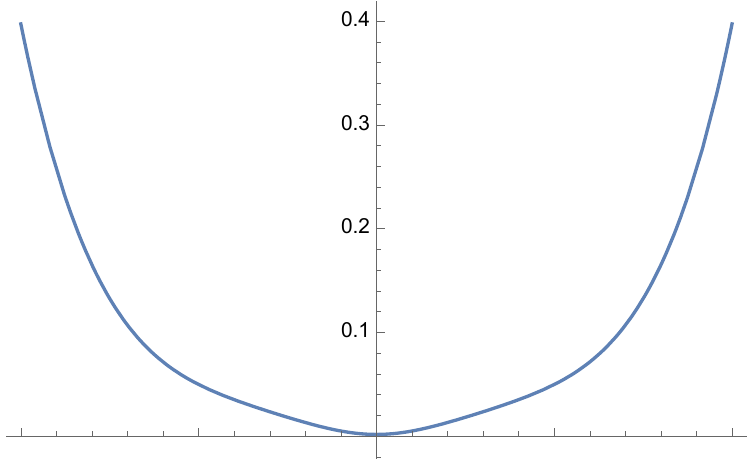} 
\includegraphics[width=0.4\linewidth]{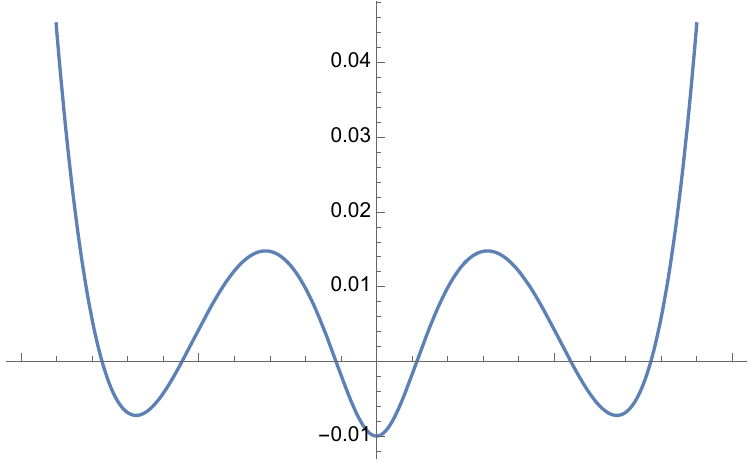} 
\includegraphics[width=0.4\linewidth]{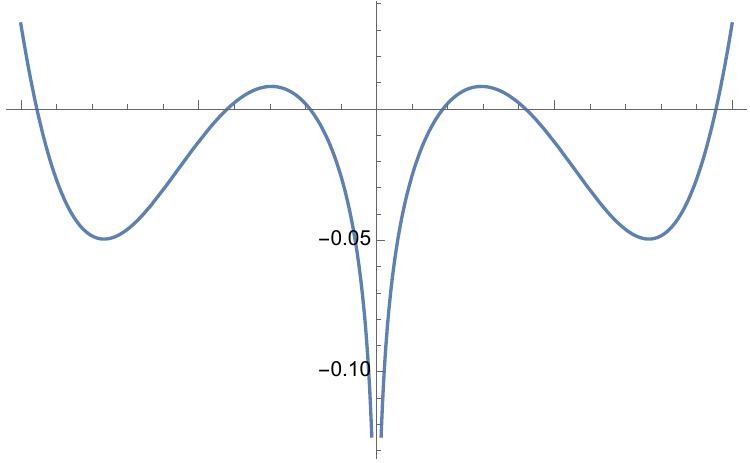} 
    \caption{1-loop effective potential in the de Sitter $d=3$ case. The bare coupling constant is $c_R=4$. The masses from left to right: principal series $m_R=10$, limit case $m_R=1$, complementary series $m_R=0.5$ and $m_R=0.001$. One observes the curious behavior of the complementary series. Further analysis is needed with the renormalization group.}
    \label{fig:1}
\end{figure}

%%%%%%%%%%%%%%%%%%%%%%%%%%%%%%%%%%%%%%%%%%
\subsection{Two-loops in $d=3$}

 The first contribution to the vacuum diagrams comes at two-loop order. There are three types of ``8-diagrams'' or ``double-tadpole'' diagrams,  say $T_a, T_b, T_c$, and two types of ``watermelon'' diagrams, say $W_1, W_2$.
\begin{itemize}
    \item 
$T_a$ is the double-tadpole with two masses $M_0$. It comes from the $\psi_0^4$ term, so it contributes with a factor 3.
\item $T_b$ is the double-tadpole with one mass $M_0$ and one mass $M_1$. It comes from the double products $\psi_0^2 \psi_j^2$ in $(\sum_{j=0}^{N-1} \psi_j^2)^2$, so it contributes with a factor $2(N-1)$.\\
\item $T_c$ is the double-tadpole with two masses $M_1$. It comes from the terms $\psi_j^4$ and the double products $\psi_i^2 \psi_j^2$, $i\neq j$ in $(\sum_{j=1}^{N-1} \psi_j^2)^2$. 
There are therefore $3(N-1)$ terms of the first type and $2 \frac {(N-1)(N-2)}2$ terms of the second type for a total factor $N^2-1$.\\
\end{itemize}
The total contribution of the double tadpoles is thus
\begin{align}
 {V}_T(\varphi_R)=\frac {c_R}{4} \Big( 3 T_3(\nu_{0})^2 +2(N-1) T_3(\nu_{0}) T_3(\nu_{1})+(N^2-1) T_3(\nu_{1})^2 \Big),
\end{align}
where $T_3(\nu)$ is given by \eqref{T3}.\\
As for the watermelons:
\begin{itemize}
    \item $W_1$ is the watermelon with three equal masses $M_0$. There are $3!=6$ of them.
\item $W_2$ is the watermelon with one mass $M_0$ and two equal masses $M_1$. There are $2(N-1)$ of them.
\end{itemize}
Therefore, the total contribution of the watermelons is
\begin{align}
 {V}_W(\varphi_R)=-\frac 12 \left(c_R \varphi_R \right)^2 \Big( 6 I_3(\nu_{0},\nu_{0},\nu_{0})+2(N-1) I_3 (\nu_{1},\nu_{1},\nu_{0})\Big),
\end{align}
where (see \cite{CEM2024}, formula (10.20))
\begin{align}
&I_{3}(x,y,w) :=- \frac {1}{32 \pi^2 (d-3) } + \frac{1-\gamma +\log (\pi )}{32 \pi ^2} +
\cr &- \sum_{\epsilon,\epsilon'=\pm} \frac{\left(\psi\left(\frac{1}{2}-i\frac w2-i\epsilon \frac x2-i\epsilon ' \frac y2\right)+\psi \left(\frac{1}{2}+i\frac w2+i\epsilon \frac x2+i\epsilon ' \frac y2\right)\right) \sinh \left(\pi  \left(w+\epsilon  x+\epsilon ' y\right)\right)}{128 \pi ^2 \sinh
   ( \pi  w) \sinh (\pi  \epsilon  x) \sinh \left(\pi  \epsilon ' y\right)}\cr&+O(d-3).
\end{align}
By introducing the function
\begin{align}
 f(x)=\psi\left(\frac {1+ix}2 \right)+\psi\left(\frac {1-ix}2 \right)
\end{align}
we have
\begin{align}
 I_3(\nu_{0},\nu_{0},\nu_{0}) +\frac {1}{32 \pi^2 (d-3) } &- \frac{1-\gamma +\log (\pi )}{32 \pi ^2}=\frac {3 f(\nu_{0})}{128 \pi^2 \sinh^2(\pi\nu_{0})} -\frac {f(3\nu_{0}) \sinh(3\pi\nu_{0})}{128 \pi^2 \sinh^3(\pi\nu_{0})}\cr
% =&\frac {3 (f(\nu_{0})-f(3\nu_{0}))}{128 \pi^2 \sinh^2(\pi\nu_{0})} -\frac {f(3\nu_{0}) }{32 \pi^2} \cr
 =&\frac {3 (f(\nu_{0})-f(3\nu_{0}))}{128 \pi^2} \coth^2 (\pi\nu_{0}) -\frac {3f(\nu_{0})+f(3\nu_{0}) }{128 \pi^2} ,
\end{align}
and
\begin{align}
 I_3(\nu_{1},\nu_{1},&\nu_{0})+ \frac {1}{32 \pi^2 (d-3) } - \frac{1-\gamma +\log (\pi )}{32 \pi ^2}=\frac {2 f(\nu_{0})}{128 \pi^2 \sinh^2(\pi\nu_{1})}\cr & -\frac {f(\nu_{0}+2\nu_{1}) \sinh(\pi(\nu_{0}+2\nu_{1}))}{128 \pi^2 \sinh^2(\pi\nu_{1}) \sinh(\pi\nu_{0})}
 +\frac {f(2\nu_{1}-\nu_{0}) \sinh(\pi(2\nu_{1}-\nu_{0}))}{128 \pi^2 \sinh^2(\pi\nu_{1}) \sinh(\pi\nu_{0})}.
\end{align}
For the reader's convenience  we summarize here  some properties of the function $f$. By construction, $f(x)=f(-x)$. By using the Gauss integral representation of the digamma function \cite{WW},
\begin{align*}
 \psi(z)=\log z-\frac 1{2z}-\int_0^\infty \left(\frac 12-\frac 1t+\frac 1{e^t-1}\right)e^{-tz}\ dt,
\end{align*}
we get the representation
\begin{align}
 f(x)=\log \frac {m^2_x}4 -\frac 2{m^2_x}-2\int_0^\infty \left(\frac 12-\frac 1t+\frac 1{e^t-1}\right)e^{-\frac t2} \cos \frac {xt}2 \ dt,
\end{align}
where $m_x^2=x^2+1$. The Binet integral representation \cite{WW}
\begin{align*}
 \psi(z)=\log z-\frac 1{2z}-2\int_0^\infty \frac {t\ dt}{(t^2+z^2)(e^{2\pi t}-1)},
\end{align*}
gives another useful representation of $f$:
\begin{align}
 f(x)=\log \frac {m^2_x}4 -\frac 2{m^2_x}-8\int_0^\infty \frac {t\ dt}{e^{2\pi t}-1} \frac {2t^2+1-\frac {m^2_x}{2}}{\left(2t^2+1-\frac {m^2_x}{2}\right)^2+m^2_x-1}.
\end{align}
From this, we get the asymptotic expansion for large $x$:\footnote{The same formulas can be obtained by the well-known asymptotic expansion for the digamma function.}
\begin{align}
    f(x)\sim& \log \frac {m_x^2}4-\frac 43 \frac 1{m_x^2},\cr
    f'(x)\sim& \frac {2x}{m_x^2}+\frac 83 \frac {x}{m_x^4}.
\end{align}
Finally, by using the identity
%\begin{align}
% \Gamma(x+iy)\Gamma(x-iy)=\Gamma(z)^2 \prod_{k=0}^\infty \left(1+\frac {y^2}{(x+k)^2} \right),
%\end{align}
\begin{align}
 \Gamma(x+iy)\Gamma(x-iy)=
 \Gamma(x)^2 \prod_{k=0}^\infty \frac{1}{\left(1+\frac {y^2}{(x+k)^2} \right)},   
\end{align}
we get 
\begin{align}
 f(x)=2\psi(1/2)+\frac {4(m_x^2-1)}{m_x^2}+\sum_{k=1}^\infty \frac {4(m_x^2-1)}{(2k+1)(m_x^2-1+(2k+1)^2)},
\end{align}
which may be useful to check the zero mass limit.\\
In all these formulas, the parameters are bare. We now need to discuss the renormalization.

%%%%%%%%%%%%%%%
\subsection{Renormalization}
Collecting all the terms, up to second order in $\hbar$, we get
\begin{align}
 {V}(\varphi_R)=& \Lambda_0+\frac{c_R^2 (N+2)}{32\pi^2}\left( \frac {1}{ (d-3) } - (1-\gamma +\log (\pi )) \right)\varphi_R^2+ \frac {m_R^2}2 \varphi_R^2 +\frac {c_R}4 \varphi_R^4\cr
 &-\frac{\nu_{0}^3}{12 \pi }-\frac{\nu_{0}^2 \log \left(1-e^{-2 \pi  \nu_{0} }\right)}{4 \pi^2}+\frac{\nu_{0} \text{Li}_2\left(e^{-2 \pi  \nu_{0} }\right)}{4 \pi^3}+\frac{\text{Li}_3\left(e^{-2 \pi  \nu_{0} }\right)}{8 \pi ^4}\cr
 &+(N-1)\left(-\frac{\nu_{1}^3}{12 \pi }-\frac{\nu_{1}^2 \log \left(1-e^{-2 \pi  \nu_{1} }\right)}{4 \pi^2}+\frac{\nu_{1} \text{Li}_2\left(e^{-2 \pi  \nu_{1} }\right)}{4 \pi^3}+\frac{\text{Li}_3\left(e^{-2 \pi  \nu_{1} }\right)}{8 \pi ^4}\right)\cr
 &+\frac {c_R}{64\pi^2} \Big( 3 \nu_{0}^2\coth^2(\pi \nu_{0}) +2(N-1) \nu_{0}\coth(\pi \nu_{0}) \nu_{1}\coth(\pi \nu_{1})\cr &+(N^2-1) \nu_{1}^2\coth^2(\pi \nu_{1}) \Big)\cr
 &-\frac {3c_R^2\varphi_R^2}{128 \pi^2} \left( 3 (f(\nu_{0})-f(3\nu_{0})) \coth^2 (\pi\nu_{0}) -3f(\nu_{0})-f(3\nu_{0}) \right)\cr
 &-\frac {c_R^2 (N-1)\varphi_R^2}{128 \pi^2} \Bigg( \frac {2 f(\nu_{0})}{ \sinh^2(\pi\nu_{1})} -\frac {f(\nu_{0}+2\nu_{1}) \sinh(\pi(\nu_{0}+2\nu_{1}))}{ \sinh^2(\pi\nu_{1}) \sinh(\pi\nu_{0})}\cr
 &+\frac {f(2\nu_{1}-\nu_{0}) \sinh(\pi(2\nu_{1}-\nu_{0}))}{ \sinh^2(\pi\nu_{1}) \sinh(\pi\nu_{0})} \Bigg).
\end{align}
Recall that the subscript $R$ denotes adimensional quantities rescaled by the de Sitter radius $R$, 
related to bare parameters via $m_R = m_0 R$, $\varphi_R = \varphi_0  \sqrt{R}$, $c_R = c_0  R$; 
consistently, $R$ is taken as the renormalized de Sitter radius.

Tadpoles being finite, divergences affect only the first mass term and are removed by mass 
renormalization alone. This soft divergence in the 3D effective potential is expected in 
super-renormalizable theories, mirroring the Minkowski case.
In $d=4$ at two loops, logarithmic divergences require a non-minimal $\xi \mathcal R \varphi^2$ coupling; in $d=3$, this would shift $m_R^2\to m_R^2+6\xi_R$. Yet, no such term is necessary as $\xi$ does not run with energy scaling. Nevertheless, we will see below that $\xi$ receives corrections from quantum fluctuations and not from renormalization: it does not run but is corrected by the quantum dynamics.

Thus, one is tempted to say that the only effective renormalization is the mass renormalization, like in the 3D Minkowskian case, and the cosmological constant. However, in de Sitter space the cosmological constant is not just an external parameter but enters the physics, and appears everywhere in the effective potential. 

The key strategy is taking 
$R$
as the renormalized de Sitter radius. Even then, finite corrections to the constant term emerge, interpretable as radiative shifts to Newton's gravitational constant.

Let us define
\begin{align}
    \delta m_0^2:=\frac{c_0^2 (N+2)}{16\pi^2}\left( \frac {1}{ (d-3) } - (1-\gamma +\log (\pi )) \right),
\end{align}
and simply proceed with a minimal subtraction scheme, by defining the renormalized mass
\begin{align}
    m^2:=m_0^2+\delta m_0^2.
\end{align}
Since all other terms are already finite, also remembering that in order to reintroduce units of measure we have to divide the potential by $R^3$ so that ${V}(\varphi):=\frac {{V}(\varphi_R)}{R^3}$, and we now call $\varphi$ the (un)renormalized field, we get for the renormalized parameters
\begin{align}
 {V}(\varphi)=& K_0+\frac {m^2}2 \varphi^2 +\frac {c}4 \varphi^4-\frac{\nu_{0}^3}{12 \pi }-\frac{\nu_{0}^2 \log \left(1-e^{-2 \pi R \nu_{0} }\right)}{4 \pi^2 R}+\frac{\nu_{0} \text{Li}_2\left(e^{-2 \pi R \nu_{0} }\right)}{4 \pi^3 R^2}+\frac{\text{Li}_3\left(e^{-2 \pi R \nu_{0} }\right)}{8 \pi ^4R^3}\cr
 &+(N-1)\left(-\frac{\nu_{1}^3}{12 \pi }-\frac{\nu_{1}^2 \log \left(1-e^{-2 \pi R \nu_{1} }\right)}{4 \pi^2 R}+\frac{\nu_{1} \text{Li}_2\left(e^{-2 \pi R \nu_{1} }\right)}{4 \pi^3R^2}+\frac{\text{Li}_3\left(e^{-2 \pi R \nu_{1} }\right)}{8 \pi ^4R^3}\right)\cr
 &+\frac {c}{64\pi^2} \Big( 3 \nu_{0}^2\coth^2(\pi R\nu_{0}) +2(N-1) \nu_{0}\coth(\pi R\nu_{0}) \nu_{1}\coth(\pi R\nu_{1})+(N^2-1) \nu_{1}^2\coth^2(\pi R\nu_{1}) \Big)\cr
 &-\frac {3c^2\varphi^2}{128 \pi^2} \Big( 3 (f(R\nu_{0})-f(3R\nu_{0})) \coth^2 (\pi R\nu_{0}) -3f(R\nu_{0})-f(3R\nu_{0}) \Big)\cr
 &-\frac {c^2 (N-1)\varphi^2}{128 \pi^2} \Bigg( \frac {2 f(R\nu_{0})}{ \sinh^2(\pi R\nu_{1})} -\frac {f(R(\nu_{0}+2\nu_{1})) \sinh(\pi R(\nu_{0}+2\nu_{1}))}{ \sinh^2(\pi R\nu_{1}) \sinh(\pi R\nu_{0})}\cr &+\frac {f(R(2\nu_{1}-\nu_{0})) \sinh(\pi R(2\nu_{1}-\nu_{0}))}{ \sinh^2(\pi R\nu_{1}) \sinh(\pi R\nu_{0})} \Bigg), \label{3.61}
\end{align}
where we have introduced the definitions
\begin{align}
    \nu_0^2&=:m^2+3c\varphi^2 -\frac{\cal R}6\equiv M^2_0-\frac{\cal R}6,\\
    \nu_1^2&=:m^2+c\varphi^2 -\frac{\cal R}6\equiv M^2_1-\frac{\cal R}6,
\end{align}
${\cal R}=\frac 6{R^2}$ being the Ricci scalar, and all parameters are renormalized. However, it is important to notice that these parameters are not the physical ones. The physical parameters are defined as follows.
%%%%%%%%%%%%%%%%%%%%%%%%%%%%
\subsubsection*{The cosmological constant}
The cosmological constant is determined by the value of ${V}(0)$. More precisely, if $G_3$ is the three-dimensional Newton's constant, we have
\begin{align}
   -\frac {\Lambda_{cosm}}{8\pi G_3}&=K_0-N\left(-\frac{\nu^3}{12 \pi }-\frac{\nu^2 \log \left(1-e^{-2 \pi R \nu }\right)}{4 \pi^2 R}+\frac{\nu \text{Li}_2\left(e^{-2 \pi R \nu }\right)}{4 \pi^3R^2}+\frac{\text{Li}_3\left(e^{-2 \pi R \nu }\right)}{8 \pi ^4R^3}\right)\cr
 &+\frac {cN(N+2)}{64\pi^2} \nu^2\coth^2(\pi R\nu)-\frac {(N+2)c^2\varphi^2}{128 \pi^2 \sinh^2(\pi R\nu)} \left( 3 f(R\nu)-f(3R\nu)\frac {\sinh(3\pi R\nu)}{\sinh(\pi R\nu)} \right),
\end{align}
where
\begin{align}
    \nu^2&=:m^2 -\frac{\cal R}6.
\end{align}
In the present paper, we choose to define a static potential, so that we can simply include in $K_0$ a counterterm which absorbs all terms leaving only the physical cosmological constant.
%%%%%%%%%%%%%%%%%%%%%%
%Although the cosmological constant has been renormalized here to be treated as a static parameter, the dependence of the loop contributions on mass and curvature suggests that, in a dynamical back-reaction context (In-In formalism), such corrections could lead to a time evolution of the cosmological constant itself.

%%%%%%%%%%%%%%%%%%%%%%%%%%%%
\subsubsection*{The physical mass}
Let us consider the case of the principal series, with a choice of the parameters such that the potential is convex (small $c$). In this case, the minimum of the potential is at $\varphi=0$.
In this situation, the physical mass is determined by the equation
\begin{align}
    m^2_{phys}=&2 \frac {d {V}}{d\phi^2}(0)=m^2-(N+2)\frac {c\nu}{4\pi}\coth (\pi R\nu)\cr 
    &+\frac{(N+2)^2c^2}{32\pi^2}\Big(\coth^2(\pi R\nu)-\pi R\nu \coth(\pi R\nu)(\coth^2(\pi R\nu))-1\Big)\cr
    &-\frac {(N+2)c^2}{64\pi^2} \Big( 3 (f(R\nu)-f(3R\nu)) \coth^2 (\pi R\nu) -3f(R\nu)-f(3R\nu) \Big).
\end{align}
This physical mass contains both the kinematical and the geometrical contribution, in the sense that it must be considered partially as a correction to the kinematical mass and part to the geometric term $\xi {\cal R}$. However, such separation is not immediate in the generic regime, while can be understood in the weak gravity limit. We will discuss it in Sec. \ref{sec:flat-limit}.

%%%%%%%%%%%%%%%%%%%%%%%%%%%%
\subsubsection*{The physical coupling}
The physical coupling is determined by the equation
\begin{align}
    c_{phys}=&2 \frac {d^2 {V}}{d(\phi^2)^2}(0)=c-\frac {(N+8)c^2}{8\pi\nu}\coth (\pi R\nu)+\frac {(N+8)c^2R}{8}\Big(\coth^2 (\pi R\nu)-1\Big) \cr &+\frac {27c^3 R}{32\pi^2 \nu} \Bigg[-\frac 12 \Big(\coth^2 (\pi R\nu)-1\Big)f'(R\nu)+\frac 12 \Big(1+3\coth^2 (\pi R\nu)\Big)f'(3R\nu) \cr
    &+\pi \Big(\coth^2 (\pi R\nu)-1\Big)\coth (\pi R\nu) \Big(f(R\nu)-f(3R\nu)\Big) \Bigg]\cr
    &+\frac {(N-1)c^3 R}{32\pi^2 \nu} \Bigg[-\frac 12 \Big(\coth^2 (\pi R\nu)-1\Big)f'(R\nu)+\frac 52 \Big(1+3\coth^2 (\pi R\nu)\Big)f'(3R\nu) \cr
    &+5\pi \Big(\coth^2 (\pi R\nu)-1\Big)\coth (\pi R\nu) \Big(f(R\nu)-f(3R\nu)\Big) \Bigg].
\end{align}
Again, we see that the relation between the coupling parameter $c$ and the physical coupling is quite cumbersome. In order to get better control of the physics, it is helpful to look at the flat limit.

%%%%%%%%%%%%%%%
\subsection{Flat limit}\label{sec:flat-limit} 
To take the flat limit, we consider $m\geq \frac 1R$, which amounts to $mR\geq 1$. Noticing that $\nu_{j R}$ grows like $R$ and dropping terms that in ${V}(\varphi):=\frac {{V}(\varphi_R)}{R^3}$ decrease exponentially or faster than $\frac 1{R^2}$, we get
\begin{align}
 {V}(\varphi)=& K_0+ \frac {m^2}2 \varphi^2 +\frac {c}4 \varphi^4-\frac{\nu_{0}^3}{12 \pi}-\frac{(N-1)\nu_{1}^3}{12 \pi}+\frac {c}{64\pi^2} \Big( 3 \nu_{0}^2 +2(N-1) \nu_{0} \nu_{1}+(N^2-1) \nu_{1}^2 \Big)\cr
 &+\frac {3c^2\varphi^2}{32 \pi^2} \left( \log \frac {1+9R^2\nu_{0}^2}4-\frac 43 \frac 1{1+9R^2\nu_{0}^2}\right)\cr
 &+\frac {c^2 (N-1)\varphi^2}{32 \pi^2} \left( \log \frac {1+R^2(\nu_{0}+2\nu_{1})^2}4-\frac 43 \frac 1{1+R^2(\nu_{R}+2\nu_{1})^2} \right)+O(R^{-4})\cr
 =& K_0+ \frac {m^2}2 \varphi^2 +\frac {c}4 \varphi^4-\frac{\nu_{0}^3}{12 \pi}-\frac{(N-1)\nu_{1}^3}{12 \pi}+\frac {c}{64\pi^2} \Big( 3 \nu_{0}^2 +2(N-1) \nu_{0} \nu_{1}+(N^2-1) \nu_{1}^2 \Big)\cr
 &+\frac {3c^2\varphi^2}{32 \pi^2} \left( \log \frac {27\nu_{0}^2}{2{\cal R}}- \frac {\cal R}{162\nu_{0}^2}\right)+\frac {c^2 (N-1)\varphi^2}{32 \pi^2} \left( \log \frac {3(\nu_{0}+2\nu_{1})^2}{2{\cal R}}- \frac {\cal R}{18(\nu_{R}+2\nu_{1})^2} \right)\cr&+O(R^{-4})\cr
 =& K_0+ \frac {m^2}2 \varphi^2 +\frac {c}4 \varphi^4-\frac {M_0^3}{12\pi}\Big(1-\frac {\cal R}{4M_0^2}\Big)-\frac {(N-1)M_1^3}{12\pi}\Big(1-\frac {\cal R}{4M_1^2}\Big)\cr 
 &+\frac {c}{64\pi^2} \Bigg( 3 M_{0}^2 +2(N-1) M_{0} M_{1}+(N^2-1) M_{1}^2- \frac {(N^2+2)M_0M_1+(N-1)(M_0^2+M_1^2)}{6}{\cal R} \Bigg)\cr
 &+\frac {3c^2\varphi^2}{32 \pi^2} \left( \log \frac {27M_{0}^2}{2{\cal R}}- \frac {14\cal R}{81M_{0}^2}\right)+\frac {c^2 (N-1)\varphi^2}{32 \pi^2} \Bigg( \log \frac {3(M_{0}+2M_{1})^2}{2{\cal R}}\cr &- \frac {(3M_0^2+3M_1^2+8M_0M_1)}{3M_0M_1(M_{0}+2M_{1})^2} {\cal R}\Bigg)+O({\cal R}^2).
\end{align}
It is clear that in the limit ${\cal R}=0$, this reproduces the two-loop effective potential with $N=1$, computed in \cite{Rajantie:1996}, after the logarithmically divergent terms are absorbed in the renormalization of the mass.
Indeed, in this approximation, the mass relation takes the form
\begin{align}
    m^2_{phys}=&=m^2-(N+2)\frac {c\nu}{4\pi}+\frac{(N+2)^2c^2}{32\pi^2}+\frac {(N+2)c^2}{16\pi^2} \Big(\log \frac {27 m^2}{2\cal R}-\frac {14}{81} \frac {\cal R}{m^2}\Big)+O({\cal R}^2).
\end{align}
Here we clearly see the physical meaning of the contributions: the first three terms determine the physical kinetic mass; the very last term gives the radiative correction to the $\xi$ parameter in the contribution $\frac 12 \xi {\cal R} \varphi^2$ (which, for simplicity, we initially assumed to be zero, but it could be equally included in $m^2$); the remaining logarithmic term looks like a Lamb-shift like correction to the mass, essential for giving the correct flat limit.\\
As for the coupling constant, we get
\begin{align}
    c_{phys}=&c-\frac {(N+8)c^2}{8\pi\nu} +\frac {(5N+22)c^3}{72\pi^2 \nu^2} \Big(  1+\frac {5}{54 \nu^2} {\cal R} \Big)+O({\cal R}^2).
\end{align}
%%%%%%%%%%%%%%%%%%%%%%%%%%%%%%%%%%%
\subsection{$\beta$-function and anomalous mass dimension}
In order to compute the anomalous mass dimension, recall that the energy scaling must be restored by replacing the coupling $c$ with $c \mu^{3-d}$, so that the renormalization of the mass is
\begin{align}
    \delta m_0^2:=\frac{c^2 \mu^{3-d} (N+2)}{16\pi^2}\left( \frac {1}{ (d-3) } - (1-\gamma +\log (\pi )) \right).
\end{align}
Therefore, the anomalous mass dimension is (for $m$ the renormalized mass)
\begin{align}
    \gamma_m=\frac {\mu}{m}\partial_\mu m=-\frac {N+2}{16\pi^2} \frac {c^2}{m^2}.
\end{align}
It is also convenient to introduce the dimensionless coupling constant $g$ defined by $c=mg$. It follows that the corresponding $\beta$-function is 
\begin{align}
    \beta_g=\frac {N+2}{16\pi^2} \frac {c^3}{m^2}=\frac {N+2}{16\pi^2} m g^3.
\end{align}
These formulas coincide with the ones in the flat case \cite{Kleinert}.

%%%%%%%%%%%%%%%
%\section{Applications (?)}

%We consider a real scalar field on Euclidean de Sitter space, which is the$d$--sphere $S^d$ of radius $a$. The goal is to construct a Wilsonianrenormalization group (RG) flow for the \emph{local effective potential}in the spirit of Wegner--Houghton, but formulated directly in finitevolume and discrete spectrum, following the approach of Shepard for afinite lattice.In particular, the one--loop effective potential on $S^d$ in a constantbackground $\phi$ has the familiar form\begin{equation}	U_{\text{1-loop}}(\phi)	\;=\;	U_{\text{bare}}(\phi)	+ \frac{1}{2V_d}\sum_{l=0}^{\infty} g_l\,	\log\!\bigl[\lambda_l + U_{\text{bare}}''(\phi) + \xi R\bigr],\label{eq:1loop-Trlog}\end{equationwhere the sum over $g_l$ implements the trace over all degenerateeigenmodes at fixed $l$.
%We now want to construct a \emph{discrete} Wegner--Houghton flow in whichthe effective potential $U_l(\phi)$ at ``scale'' $l$ is obtained byintegrating out all modes with angular momentum $l'>l$. The resultingrecursion relation will be written in terms of the degeneracies $g_l$,or equivalently in terms of the \emph{resummed} (cumulative) degeneracy$N_l=\sum_{m=0}^{l}g_m$.

\section{Anti de Sitter effective potentials}
\label{4}
We compute the two-loop effective potential for the analogous model on AdS 3-dimensional spacetime, see \cite{bemgen} for field quantization on AdS. In this case, in place of the dimensional regularization, we will use point-splitting regularization.
\subsection{The one loop case}
The $d$-dimensional real AdS spacetime
with radius $R>0$ may be visualized as the manifold
\begin{align}
    AdS_d = \{x \in \bR^{d+1}\ :\ x^2= x\cdot x = R^2\}
\label{s.1}
\end{align}
where the scalar product $x\cdot x$ is intended in the sense of the ambient space $ \bR^{d+1}$  with two timelike directions and metric mostly minus as follows:
\begin{align}
   x\cdot y  = x^0 y^0-x^1y^1-\ldots-x^{d-1}y^{d-1} +x^dy^d. 
\label{metricads}\end{align}
%The vector $e_\mu \in \bR^{d+1}$ has coordinates$e_\mu^\nu = \delta_{\mu \nu}$.
The complexification of the AdS manifold is defined analogously 
\beq
AdS_d^{(c)} = \{z =x+iy \in \bC^{d+1}\ :\ z^2 = R^2\};
\label{complexmanifold}\endq
 $z\in AdS^{(c)}$ if and only if  $x^2 -y^2=R^2$ and $x\cdot y = 0$, i.e. the real and imaginary parts of $z$ are orthogonal w.r.t. the scalar product in the ambient space. 
 
 The maximally analytic two-point function has the following expression of a massive scalar field (which is in general well-defined  only on the universal covering of the AdS manifold)
\begin{align}
&W^{(AdS)}_\nu(z_1,z_2)=
{1 \over (2\pi)^{d\over 2}} (\zeta^2-1)^{-\frac{d-2}4}
e^{-i\pi{d-2\over 2}} Q_{-{1\over 2}+\nu}^{d-2\over 2}(\zeta)
\label{s.9}=\\ & =\frac{ \Gamma
   \left(\frac{d-1}{2}+\nu \right)  \,
  }{ 2\pi ^{\frac{d-1}{2}}  (2\zeta)^{ \frac{d-1}{2}+\nu }\Gamma (\nu +1)}  \ {}_2F_1\left(\frac{d-1}{4}+\frac{\nu }{2},\frac{d+1}{4}+\frac{\nu }{2};\nu
   +1;\frac{1}{\zeta^2}\right)
\label{kgtp}\end{align}
where 
\begin{equation}
    \zeta = \frac{z_1\cdot z_2}{R^2}.
\end{equation}
The Schwinger function (otherwise called the Euclidean  propagator) is the restriction of the maximally analytic  two-point function  to the Euclidean Lobachevsky manifold. Choosing the points in Eq. (\ref{s.9}) as follows 
\begin{align}
& z_0= \left(
1 ,
0 ,
\ldots
0,
0 
\right),
 \ \ \ \ \ z(u,\omega)= \left(
u,
\omega^1 \sqrt{u^2-1},
\ldots,
\omega^{d-1} \sqrt{u^2-1},
i \omega^{d} \sqrt{u^2-1}
\right), \ \ \  u>1 \label{uuu}
\end{align}
so that $ \zeta=  z_0\cdot 
z (u,\omega) = u>1$, we write  the propagator as 
\begin{equation}
 G^{(AdS)d}_{\nu}(z_0\cdot 
z (u,\omega))= G^{(AdS)d}_{\nu}(u) = G^d_{\nu}(u)=w_\nu(u) = \frac
{e^{-i\pi\frac {d-2}2}}{(2\pi)^{\frac{d}2}} (u^2-1)^{-\frac
{d-2}4} Q^{\frac {d-2}2}_{-\frac 1 2+\nu}(u). \label{kgtps}
\end{equation}
where the various parameters are related as follows:
\begin{equation}
     m^2 = \nu^2 -\frac{(d-1)^2}4.
\end{equation}
The tadpole is computed from maximally analytic two-point function by taking the limit $z_1\cdot z_2\to 1$ in the above formula and, as  before, is finite for $d<2$:
\begin{equation}
 T_d(\nu)=   \frac{2^{-d} \pi ^{-d/2} \Gamma \left(1-\frac{d}{2}\right) \Gamma \left(\frac{d-1}{2}+\nu
   \right)}{\Gamma \left(\frac{3-d}{2}+\nu\right)}
   \end{equation}
The  meromorphic  continuation of this formula provides a dimension-regularized integral representation of the 1-loop correction to the bare potential: 
\begin{eqnarray}
V_0(\varphi_R)=2^{-d} \pi ^{-d/2} \Gamma \left(1-\frac{d}{2}\right)\int^{\nu(\phic_R)}    \frac{ \Gamma \left(\frac{d-1}{2}+
    x\right)}{\Gamma
   \left(\frac{3-d}{2}+ x\right) }  x dx \label{potads}
\end{eqnarray}
where 
\begin{equation}
    \nu(\phic_R)=\pm \left(m_R^2+3c_R \phic_R^2+\frac{(d-1)^2}{4}\right)^{\frac 12}, \ \ \ \nu>-1.
\end{equation}
The restriction $\nu >-1$ expresses the famous Breitenlohner and Freedman bound \cite{Breitenlohner,bertola}. 
In dimensional regularization, the above formula directly gives the 1-loop correction to the effective potential in odd dimensions, much simpler than in de Sitter case; for instance at $d=1$ and $d=3$ 
\begin{equation}
T_1(\nu) = \frac{1}{2 
\nu}, \ \ T_3(\nu) = -\frac{\nu}{4\pi} 
\end{equation}
so that 
  \begin{eqnarray}
   && V^{(1)}_0(\phic_R)=  \frac{1}{2} \left(\sqrt{m_R^2+3 c_R \phic_R^2 }-m\right) \\
    && V^{(3)}_0(\phic_R) = \frac{{\left(m_R^2+1\right)^{3/2}}-\left(m_R^2+3 c_R \phic_R ^2{+1}\right)^{3/2}}{12 \pi }.
   \end{eqnarray}
This is the flat space result! 
While this exact coincidence happens only in $d=1$ and $d=3$ (with a constant shift)  it remains essentially verified in any odd dimension in the sense that  when $d=2n+1$ the dominant term computed with the help of Eq. (\ref{potads}) is identical to the flat space result (\ref{225}).\\
In even dimensions the result is more complicate; for $d=2$ it reproduces the results in \cite{Sakai}.
%%%%%%%%%%%%%%%%%%%%%%%%%%%%%%%%%%
\subsection{Two-loop at $d=3$}
We need to compute the banana diagram with three independent masses, \cite{HSUAdS}.
The  $n$-loop banana integral on the Lobachevsky Euclidean manifold  with $n+1$ edges and two vertices:
\begin{eqnarray}
I_{n+1}(\nu_1,\ldots,\nu_{n+1},d)= \int_{H_d} G^{d}_{\nu_1}(x\cdot z)  G^{d}_{\nu_2}(x\cdot z) \ldots  G^{d}_{\nu_{n+1}}(x\cdot z )\sqrt{g(z)}\, dz ,  \label{diagram}\end{eqnarray}
where $y$ varies on  $H_d$ and $x$ is a fixed reference point. The above definition has to be intended as a  regularization of an expression that in general is divergent. 
Using the coordinates (\ref{uuu}) with $u=\cosh v$ and integrating over the angles  (\ref{diagram}) reduces to 
\begin{eqnarray}
I_{n+1}(\nu_1,\ldots,\nu_{n+1},d)=
  \frac {2\pi ^{\frac {d}{2}}}{\Gamma \left({\frac {d}{2}}\right)} \int _0^\infty G^{d}_{\nu_1}(\cosh v )  G^{d}_{\nu_2}(\cosh v  )  \ldots G^{d}_{\nu_{n+1}}(\cosh v )(\sinh v)^{d-1}dv.\label{bub}
\end{eqnarray}
%The zero-loop case gives \begin{equation}I_1(\nu,d)   = \frac 1{\nu^2-\frac {(d-1)^2}4}= \frac 1{m^2}. \label{I1}\end{equation}
We want to compute the two-loop diagram at $ d=3$. 
\begin{eqnarray}
I_{3}(\nu_1,\nu_2,\nu_{3},d)=
4\pi \int _0^\infty G^{3}_{\nu_1}(\cosh v )  G^{3}_{\nu_2}(\cosh v  )  G^{3}_{\nu_{3}}(\cosh v )(\sinh v)^{2}dv.\label{bub}
\end{eqnarray}
Here the propagator reduces to an elementary function: 
\cite[Eq. 12, p 150]{bateman}:
\begin{align}
& G^3_\nu(\cosh v) = {e^{-{i\pi \over 2}} Q_{\nu-{1\over 2}}^{1\over 2}(\cosh v)  \over 2\pi \sqrt{2 \pi \sinh v}} 
= \frac {e^{-\nu v}}{4 \pi \sinh  v}  .
\end{align}
Notice that for each choice of mass one can take two values of $\nu$, but, for negative $\nu$, only $\nu>-1$ gives a good behaviour at infinity, confirming the Breitenlohner-Freedman (BF) analysis \cite{Breitenlohner,bertola}.
Indeed, for large $v$, we see that $G$ goes like $\frac 1{2\pi}u^{-\Delta}$, where $\Delta=\frac {d-1}2+\nu$ is the conformal weight of the dual theory. The choice of the sign of $\nu$ thus determine a quantum theory dual to a different conformal field theory. For the negative $\nu$, however, the resulting quantum field theory is unitary only for $-1<\nu<0$.
\\
The bubble with two independent  mass parameters $\nu_1$ and $\nu_2$ is readily decomposed into its K\"all\'en-Lehmann series  by an elementary manipulation:
\begin{align}
G^3_{\nu_1}(\cosh v) G^3_{\nu_2}(\cosh v)& = \frac {e^{-({\nu_1}+{\nu_2}) v}}{16 \pi^2 \sinh^2  v}% = \frac {e^{-({\nu_1}+\nu_2+1) v}}{8 \pi^2 \sinh  v(1-e^{-2v}) } 
= \sum_{k=0}^\infty \frac {e^{-({\nu_1}+\nu_2+1+2k ) v}}{8 \pi^2 \sinh  v }   = \frac{1}{2\pi} \sum^\infty_{k=0} G^3_{2k+1+{\nu_1}+{\nu_2}}(\cosh v). 
\end{align}
so that 
\begin{eqnarray}
I_{3,K}(\nu_1,\nu_2,\nu_{3},d)=%2 \sum^\infty_{k=0} \int _1^\infty   G^3_{2k+1+\nu_1+\nu_2}(\cosh v)    G^{3}_{\nu_{3}}(\cosh v )(\sinh v)^{2}dv \cr 
\frac{1}{8\pi^2} \sum^\infty_{k=0} \int _K^\infty   e^{-(\nu_1+\nu_2+\nu_3+1+2k ) v} dv. \cr 
\end{eqnarray}
In the last step we exchanged the integral and the series and put a UV cutoff in the integration domain. Notice that in the standard quantization, corresponding to positive $\nu_j$ and with no restriction to the masses, there are no problems of convergence at infinity. But for small masses, in the BF interval, one can choose $\nu<0$. In this case, the convergence of the integral requires
\begin{equation}
    \nu_1+\nu_2+\nu_3+1>0. \label{BF-condition}
\end{equation}
This condition strictly limits the values of $\varphi$ such that the calculation of the effective potential makes sense. \\
The above expression is readily computed and provides the regularized 2-loop bubble:
\begin{eqnarray}
I_{3,K}(\nu_1,\nu_2,\nu_{3},d)&=&%2 \sum^\infty_{k=0} \int _1^\infty   G^3_{2k+1+\nu_1+\nu_2}(\cosh v)    G^{3}_{\nu_{3}}(\cosh v )(\sinh v)^{2}dv \cr 
\frac{e^{-K (\nu_1+\nu_2+\nu_{3} +1)} \, _2F_1\left(1,\frac{\nu_1+\nu_2+\nu_{3} +1}{2};\frac{\nu_1+\nu_2+\nu_{3} +3}{2};e^{-2
   K}\right)}{8 \pi ^2 (\nu_1+\nu_2+\nu_{3} +1)} \cr &\simeq& -\frac{1 }{16 \pi ^2} \psi \left(\frac{\nu_1+\nu_2+\nu_{3} +1}{2}\right)-\frac{\log (2 K)+\gamma }{16 \pi ^2}.
\end{eqnarray}
This is essentially the point-splitting renormalization scheme. In order to renormalize the theory up to two loops, we need to work in the same renormalization scheme at one loop. In the point-splitting renormalization scheme, with point separation $K$, the tadpole in $d=3$ is
\begin{align}
T_3(\nu)= G^3_\nu(\cosh K) = \frac {e^{-\nu K}}{4 \pi \sinh  K} \simeq \frac 1{4\pi K}-\frac {\nu}{4\pi}.
\end{align}
If $M(\varphi_R)$ is the field-dependent effective mass and $\nu(\varphi_R)=\sqrt{M(\varphi_R)^2+1}$, the one loop contribution to the effective potential is therefore
\begin{align}
    V^{(3)}_{\rm 1loop}(\varphi_R) = \frac {\nu(0)^3-\nu(\varphi_R)^3}{12 \pi}+\frac {M(\varphi_R)^2-M(0)^2}{8\pi K}.
\end{align}
We are now ready to write the point-splitting regularized effective potential up to two loops:
\begin{align}
{V}(\varphi_R)=& \Lambda_0+ \frac {m_R^2}2 \varphi_R^2 +\frac {c_R}4 \varphi_R^4+\hbar\frac {N+2}{8\pi K} c_R \varphi_R^2
 +\hbar\frac {N\left(m_R^2+1\right)^{\frac 32}-\nu_0(\varphi_R)^3-(N-1)\nu_1(\varphi_R)^3}{12 \pi}\cr
 &+\hbar^2\frac {c_R}{4} \Big[ 3 \Big(\frac 1{4\pi K}-\frac {\nu_0(\varphi_R)}{4\pi}\Big)^2 +2(N-1) \Big(\frac 1{4\pi K}-\frac {\nu_0(\varphi_R)}{4\pi}\Big) \Big(\frac 1{4\pi K}-\frac {\nu_1(\varphi_R)}{4\pi}\Big)\cr
 &+(N^2-1) \Big(\frac 1{4\pi K}-\frac {\nu_1(\varphi_R)}{4\pi}\Big)^2 \Big] + \hbar^2(N+2)\frac {c_R^2 \varphi_R^2}{16\pi^2} (\log (2 K)+\gamma)\cr
 &+\hbar^2 \frac {c_R^2 \varphi_R^2}{16\pi^2} \left( 3 \psi \left(\frac{3\nu_0(\varphi_R) +1}{2}\right)+(N-1) \psi \left(\frac{\nu_0(\varphi_R)+2\nu_1(\varphi_R) +1}{2}\right)\right)+O(\hbar^3), \label{Vads}
\end{align}
where we have restored $\hbar$ for practical purposes. 
We see that in this scheme the divergences appear in a bit more complicated way than in dimensional regularization. Like in the $dS$ case, we must divide ${V}_{\rm 2loop}(\varphi_R)$ by $R^3$ and then define the renormalized parameters. In order to simplify the notations, we assume momentarily $R=1$, and will then restore it at the end. Next, we introduce the renormalized field $\varphi_r$, mass $m_r$ and coupling constant $c_r$, so that
\begin{align}
    \varphi_R=\varphi_r\sqrt{Z_\phi}, \qquad m_R^2=m_r^2 \frac {Z_{m^2}}{Z_{\phi}}, \qquad c_R=c\frac {Z_c}{Z_\phi^2}.
\end{align}
By setting\footnote{for $Z_\phi$ we can take only $Z_\phi'$, since $Z_\phi''$ enters at the third order in $\hbar$} 
\begin{align}
    Z_a=&1+\hbar Z_a'+\hbar^2 Z_a''+O(\hbar^3),\\
    \Lambda_0=&\Lambda+\hbar \Lambda'+\hbar^2 \Lambda'',
\end{align}
replacing in \eqref{Vads}, and expanding up to order $\hbar^2$ (and finally setting $\hbar=1$), we get from the renormalization conditions
\begin{align}
   {V}(\varphi_r)=& \Lambda+ \frac {m_r^2}2 \varphi_r^2 +\frac {c_r}4 \varphi_r^4
 +\frac {N\left(m_r^2+1\right)^{\frac 32}-\nu_0(\varphi_r)^3-(N-1)\nu_1(\varphi_r)^3}{12 \pi}\cr
 &+\frac {c_r}{64\pi^2} \Big[ 3 \nu_0(\varphi_r)^2 +2(N-1) \nu_0(\varphi_r) \nu_1(\varphi_r)+(N^2-1) \nu_1(\varphi_r)^2 \Big]\cr
 &+ \frac {c_r^2 \varphi_r^2}{16\pi^2} \left( 3 \psi \left(\frac{3\nu_0(\varphi_r) +1}{2}\right)+(N-1) \psi \left(\frac{\nu_0(\varphi_r)+2\nu_1(\varphi_r) +1}{2}\right)\right), \label{Vrads} 
\end{align}
and 
\begin{align}
    Z_\phi=& Z_c=1, \\
    Z_{m^2}'=& -\frac {N+2}{4\pi K} \frac {c_r}{m_r^2}, \qquad Z_{m^2}''=-\frac {N+2}{8\pi^2}\frac {c_r^2}{m_r^2}(\log (2K)+\gamma), \\
    \Lambda'=&0, \qquad \Lambda''=\frac {N(N+2)c_r}{32\pi^2 K} \left(\frac 1{2 K}-\left(\frac 1{R^2}+m_r^2\right)^{\frac 12}\right),
\end{align}
where 
\begin{align}
    \nu_0(\varphi_r)=&\sqrt{m_r^2+3c_r \varphi_r^2+\frac 1{R^2}}, \\
    \nu_1(\varphi_r)=&\sqrt{m_r^2+c_r \varphi_r^2+\frac 1{R^2}},
\end{align}
and we have restored $R$. Like in the de Sitter case, we see that we have only mass renormalization:
\begin{align}
    m^2=m_r^2-\frac {N+2}{4\pi K} c_r-\frac {N+2}{8\pi^2} c_r^2(\log (2K/R)+\gamma).
\end{align}
Notice that the above formula for the potential is also valid in the BF region
\begin{align}
    \nu_0(\varphi_r)=&-\sqrt{m_r^2+3c_r \varphi_r^2+\frac 1{R^2}}, \\
    \nu_1(\varphi_r)=&-\sqrt{m_r^2+c_r \varphi_r^2+\frac 1{R^2}},
\end{align}
for which, however, accordingly with \eqref{BF-condition}, we must require the condition
\begin{eqnarray}
    3\nu_0(\varphi_r)>-1
\end{eqnarray}
to avoid the poles in the digamma functions $\psi$. Thus, the effective potential in the BF region looks to be just the analytic continuation in $\nu$ of the one in standard quantization, along the real axis, subject to the condition \eqref{BF-condition}. However, of course, the two results are not related since they correspond to two different quantum field theories. Curiously, this bound is only manifest in the perturbative expansion of the effective potential from the two-loop order onward.  

%%%%%%%%%%%%%%%%%%%%%%%%%%%%%%%%%%%
\subsection{$\beta$-function and anomalous mass dimension}
After recalling that the energy scale is given by $\mu=1/K$, we can absorb the local counterterms, which do not contribute to the anomalous dimension, in the regular part of the mass, and write
\begin{align}
    m^2_{(reg)}=m_r^2+\frac {N+2}{8\pi^2} c_r^2(\log (2\mu R)+\gamma).
\end{align}
The mass anomalous dimension is therefore
\begin{align}
    \gamma_m=-\frac {N+2}{16\pi^2} \frac {c_r^2}{m_r^2}.
\end{align}
This is exactly the same result as in the flat case. It is convenient also to introduce a dimensionless coupling constant $g_r$ through the expression
\begin{align}
    c_r=m_r g_r.
\end{align}
The associated beta function is therefore
\begin{align}
    \beta_g=-c_r \gamma_m=\frac {N+2}{16\pi^2} \frac {c_r^3}{m_r^2}.
\end{align}
Again, this is the same result as in the flat case. Nevertheless, we remark that, in both dS and AdS cases, what makes the difference is the relation among the renormalized parameters $m_r, g_r$ and the physical parameters $m_p, g_p$.

%%%%%%%%%%%%%%%%%%%%%%%%%%%%%%%
\section{Conclusions}

In this work we have derived the first systematic treatment of one-loop effective potentials for interacting scalar fields in curved spacetimes, with explicit results for maximally symmetric de Sitter and anti-de Sitter backgrounds. Our key results can be summarized as follows:

We determined a \emph{general formula} (Eq.\eqref{LS2}) for the one-loop effective potential in arbitrary curved geometries that satisfies the simple differential condition (\ref{conditio}) for the propagator. Next, we have specified our formula to the de Sitter and anti de Sitter maximally symmetric spaces. In particular, for de Sitter, we compute the effective potential for the scalar theory $\varphi^4$ with symmetry $O(N)$ for any dimension, with an emphasis on dimensions $1,2,3,$ and 4. This is done in dimensional regularization. For the principal series, the potential is convex for small interaction parameter as compared to the mass parameter. Convexity is lost for large values of the coupling constant. This is not surprising, since the perturbative approach is expected to become inefficient as the coupling increases. In principle, convexity could be recovered by including higher-order corrections or nonperturbative methods. A possible approach to recover convexity is proposed, for example, in \cite{Wiedemann}. However, that method requires infinite volume, while we are working on a sphere of finite volume. In dimension $d=3$, still in dimensional regularization, we have extended the calculations to two loops, and computed the $\beta$-function and the anomalous mass dimension. The final expressions are identical to the ones for the flat case but with different relations between the renormalized parameters and the physical parameters. We also performed the flat limit $R\to\infty$, recovering the standard Minkowski results, confirming the construction's consistency and enabling precision comparisons between curved/flat physics.

Next, we have computed the two-loop effective potential for the same model on $AdS_3$. To illustrate the power of our methods in the configuration space, now, we worked with the point-splitting regularization. This resulted in a very simple calculation also at two loops. The main difference is that non-logarithmic divergent terms appear already at one loop. Since they are of local type, these do not contribute to the $\beta$-function and anomalous mass dimension, and we find once more the same expression as in the flat case. These results also extend to the Breitenlohner-Freedman region, with a lower bound on $\nu$ that is manifest only beyond one loop.

These results have immediate implications across multiple frontiers. Our $d=4$ de Sitter effective potentials (Eq. \eqref{3.30}) provide a reliable framework for Higgs stability and radiative symmetry breaking during inflation, where flat space approximations fail catastrophically.
The $d=3$ results (Eqs. \eqref{3.61}, \eqref{Vrads}) are directly applicable to critical phenomena modeled by QFTs on $S^3$ and $H^3$. These could be suitably analyzed by using non-perturbative functional renormalization group methods. These application will be considered in future work

\section*{Acknowledgments}
We thank Riccardo Guida for the valuable discussions and explanations and for having brought the reference \cite{Wiedemann} to our attention.


\begin{thebibliography}{99}


%%%%%%%%%%%%%%%%%%%%%%%%%%%%

\bibitem{Sher}
M.~Sher,
``Electroweak Higgs Potentials and Vacuum Stability,''
Phys. Rept. \textbf{179} (1989), 273-418

\bibitem{Branchina}
V.~Branchina, H.~Faivre and V.~Pangon,
``Effective potential and vacuum stability,''
J. Phys. G \textbf{36} (2009), 015006

\bibitem{ColemanW}
S. ~Coleman and E.~Weinberg, ``{Radiative corrections as the origin of spontaneous symmetry breaking},''
  {\em Phys. Rev.} {\bf D7}, 1888 (1973).

\bibitem{Guida}
R.~Guida and J.~Zinn-Justin,
``Critical exponents of the N vector model,''
J. Phys. A \textbf{31} (1998), 8103-8121

\bibitem{Kleinert}
H.~Kleinert and V. Schulte-Frohlinde,
``Critical Properties of $\Phi^4$ theories,''
World Scientific, Berlin (2001)
\bibitem{Zinn-Justin:2002ecy}
J.~Zinn-Justin,
``Quantum field theory and critical phenomena,''
Int. Ser. Monogr. Phys. \textbf{113} (2002), 1-1054


\bibitem{Isidori}
G.~Isidori, G.~Ridolfi and A.~Strumia,
``On the metastability of the standard model vacuum,''
Nucl. Phys. B \textbf{609} (2001), 387-409

\bibitem{Fairbairn}
M.~Fairbairn, P.~Grothaus and R.~Hogan,
``The Problem with False Vacuum Higgs Inflation,''
JCAP \textbf{06} (2014), 039

\bibitem{Anderson}
P. R. Anderson and R. Holman, 
``Effective potential for the O(N)-symmetric model in static homogeneous spacetimes,''
Phys. Rev. D 34 (1986) 2277

\bibitem{Elizalde}
E. Elizalde, K. Kirsten and S. D. Odintsov, 
``Effective Lagrangian and the back-reaction problem in a self-interacting $O(N)$ scalar theory in curved spacetime,''
Phys. Rev. D 50 (1994) 5137

\bibitem{Buchbinder}
I. L. Buchbinder, S. D. Odintsov and I. L. Shapiro, ``Effective Action in Quantum Gravity,'' IOP, Bristol, 1992.

\bibitem{Guth}
A.~H.~Guth,
``Inflationary universe: A possible solution to the horizon and flatness problems,''
Phys. Rev. D Vol. \textbf{23}, 2 (1981), 347--356

\bibitem{Wadscft}
E.~Witten,
``Anti de Sitter space and Holography,''
Adv.Theor.Math.Phys.2:253-291,1998

\bibitem{bgm}
J.~Bros, U.~Moschella and J.~P.~Gazeau,
``Quantum field theory in the de Sitter universe,''
Phys. Rev. Lett. \textbf{73} (1994), 1746-1749

\bibitem{bm}
J.~Bros and U.~Moschella,
``Two point functions and quantum fields in de Sitter universe,''
Rev. Math. Phys. \textbf{8} (1996), 327-392

\bibitem{bemgen}
J.~Bros, H.~Epstein and U.~Moschella,
``Towards a general theory of quantized fields on the anti-de Sitter space-time,''
Commun. Math. Phys. \textbf{231} (2002), 481-528

\bibitem{umads}
U.~Moschella,
``Anti-de Sitter, plane waves and quantum field theory,''
Phys. Lett. B \textbf{871} (2025), 139979

\bibitem{Sciaccaluga}
S.~Y.~Lee and A.~M.~Sciaccaluga,
``Evaluation of Higher Order Effective Potentials with Dimensional Regularization,''
Nucl. Phys. B \textbf{96} (1975), 435-444


\bibitem{CEM2024}
S.~L.~Cacciatori, H.~Epstein and U.~Moschella,
``Loops in de Sitter space,''
JHEP \textbf{07} (2024), 182

\bibitem{bemadskl}
J.~Bros, H.~Epstein, M.~Gaudin, U.~Moschella and V.~Pasquier,
``Anti de Sitter quantum field theory and a new class of hypergeometric identities,''
Commun. Math. Phys. \textbf{309} (2012), 255-291

\bibitem{HSUAdS} S.~L.~Cacciatori, H.~Epstein and U.~Moschella,
``Loops in anti de Sitter space,''
JHEP \textbf{08} (2024), 109. 

\bibitem{Rajantie:1996}
A.~K.~Rajantie,
``Feynman diagrams to three loops in three-dimensional field theory,''
Nucl. Phys. B \textbf{480} (1996), 729-752
[erratum: Nucl. Phys. B \textbf{513} (1998), 761-762]

\bibitem{Inami}
T.~Inami and H.~Ooguri,
``One Loop Effective Potential in Anti-de Sitter Space,''
Prog. Theor. Phys. \textbf{73} (1985), 1051

\bibitem{Tachyons1}
J.~Bros, H.~Epstein and U.~Moschella,
``Scalar tachyons in the de Sitter universe,''
Lett. Math. Phys. \textbf{93} (2010), 203-211

\bibitem{Tachyons2}
H.~Epstein and U.~Moschella,
``de Sitter tachyons and related topics,''
Commun. Math. Phys. \textbf{336} (2015) no.1, 381-430

\bibitem{bateman} A. Erd\'elyi,  (Ed.)   
{`` The Bateman project:  Higher Transcendental Functions,''}  
Vol.I.  New York:  McGraw-Hill Book Company (1953)

\bibitem{VicenteGarcia-Consuegra:2025two}
L.~Vicente Garc{\'\i}a-Consuegra and A.~Rajantie,
``Scalar field effective potentials in de Sitter spacetime,''
[arXiv:2511.23076 [hep-th]].

\bibitem{Fujimoto}
Y.~Fujimoto and H.~Ishihara,
``One Loop Effective Potential for Lambda $\Phi^4$ Theory in De Sitter Space,''
Phys. Lett. B \textbf{191} (1987), 46-50

\bibitem{Shore}
G.~M.~Shore,
``Radiatively Induced Spontaneous Symmetry Breaking and Phase Transitions in Curved Space-Time,''
Annals Phys. \textbf{128} (1980), 376

\bibitem{Esposito}
G.~Esposito, G.~Miele and L.~Rosa,
``One loop effective potential for SO(10) GUT theories in de Sitter space,''
Class. Quant. Grav. \textbf{11} (1994), 2031-2044

\bibitem{Weinberg} S.~Weinberg ``The Quantum Theory of Fields,'' Cambridge University Press (1996), Vol II.

\bibitem{WW}
E.T.~Whittaker, G.N.~Watson, ``A Course of Modern Analysis,'' 5th ed. Moll VH, ed. Cambridge University Press (2021)

\bibitem{Sakai}
N.~Sakai and Y.~Tanii,
``Effective Potential in Two-dimensional Anti-de Sitter Space,''
Nucl. Phys. B \textbf{255} (1985), 401

\bibitem{Breitenlohner}
P.~Breitenlohner and D.~Z.~Freedman,
``Stability in Gauged Extended Supergravity,''
Annals Phys. \textbf{144} (1982), 249
\bibitem{bertola}
M.~Bertola, J.~Bros, V.~Gorini, U.~Moschella and R.~Schaeffer,
%``Decomposing quantum fields on branes,''
Nucl. Phys. B \textbf{581} (2000), 575-603
doi:10.1016/S0550-3213(00)00280-7
[arXiv:hep-th/0003098 [hep-th]].
\bibitem{Wiedemann}
Urs Achim Wiedemann
``Non-differentiability of the effective potential,''
Nuclear Physics B, Volume 406, Issue 3, 4 October 1993, Pages 808-822



\end{thebibliography}
\end{document}